\documentclass[11pt]{article}
\usepackage[hmargin=1in,vmargin=1in]{geometry}

\usepackage{multirow}
\usepackage[utf8]{inputenc}
\usepackage{amsmath,amssymb,dsfont}
\usepackage{calc}
\usepackage{amsthm}
\usepackage{float}

\floatstyle{boxed}
\newfloat{constr}{htb!}{lop}
\floatname{constr}{Construction}

\long\def\symbolfootnote[#1]#2{\begingroup%
  \def\thefootnote{\fnsymbol{footnote}}\footnote[#1]{#2}\endgroup}

\date{}
\author{Mahdi Cheraghchi\thanks{Computer Science Department, Carnegie Mellon University, Pittsburgh, PA 15213, USA.
Email: \texttt{cheraghchi@cmu.edu}. Part of work was done while the author was with the
School of Computer and Communication Sciences, Ecole Polytechnique F\'ed\'erale de Lausanne (EPFL),
Switzerland. Research was supported in part
by the ERC Advanced investigator grant 228021 of A.~Shokrollahi, and
the Swiss National Science Foundation research grant PBELP2-133367. A preliminary
      summary of this work appears (under the same title) in
      proceedings of the $37$th International Colloquium on Automata, Languages and Programming
(ICALP~2010), Bordeaux, France, July 2010 \cite{ref:Che10}.}}  

\title{Improved Constructions for Non-adaptive Threshold~Group~Testing}

\newcommand{\eps}{\epsilon} \newcommand{\U}{\mathcal{U}}

\newcommand{\C}{\mathcal{C}}
\newcommand{\cL}{\mathcal{L}}

\newcommand{\cX}{\mathcal{X}}
\newcommand{\cY}{\mathcal{Y}}
\newcommand{\cZ}{\mathcal{Z}}
\newcommand{\tn}{{\tilde{n}}}

\newcommand{\td}{{\tilde{d}}}
\newcommand{\tk}{k}
\newcommand{\tl}{{\tilde{\ell}}}
\newcommand{\tee}{t}

\newcommand{\F}{\mathds{F}}

\renewcommand{\U}{\mathcal{U}}
\newcommand{\supp}{\mathsf{supp}}
\newcommand{\qpoly}{\mathsf{quasipoly}}
\newcommand{\wgt}{\mathsf{wgt}}
\newcommand{\dist}{\mathsf{dist}}
\newcommand{\poly}{\mathsf{poly}}
\newcommand{\zo}{\{0,1\}}
\newcommand{\cM}{M}
\newcommand{\N}{\mathds{N}}

\newcommand{\rv}[1]{\ensuremath \boldsymbol{#1}}
\newcommand{\rep}{\odot}

\newtheorem{thm}{Theorem}
\newtheorem{coro}[thm]{Corollary}
\newtheorem{lem}[thm]{Lemma}
\newtheorem{prop}[thm]{Proposition}

\theoremstyle{definition}

\newtheorem{defn}[thm]{Definition}

\newenvironment{Proof}{\begin{proof}}{\end{proof}}

\usepackage{amsrefs}

\begin{document}
\maketitle

\begin{abstract}
  The basic goal in combinatorial group testing is to identify a set
  of up to $d$ defective items within a large population of size $n
  \gg d$ using a pooling strategy. Namely, the items can be grouped
  together in pools, and a single measurement would reveal whether
  there are one or more defectives in the pool. The threshold model is
  a generalization of this idea where a measurement returns positive
  if the number of defectives in the pool reaches a fixed threshold
  $u > 0$, negative if this number is no more than a fixed lower threshold $\ell
  < u$, and may behave arbitrarily otherwise.  We study
  non-adaptive threshold group testing (in a possibly noisy setting)
  and show that, for this problem, $O(d^{g+2} (\log d) \log(n/d))$
  measurements (where $g := u-\ell-1$ and $u$ is any fixed constant) suffice to identify the
  defectives, and also present almost matching lower bounds. This
  significantly improves the previously known (non-constructive) upper
  bound $O(d^{u+1} \log(n/d))$. Moreover, we obtain a framework for
  explicit construction of measurement schemes using lossless
  condensers. The number of measurements resulting from this scheme is
  ideally bounded by $O(d^{g+3} (\log d) \log n)$.  Using
  state-of-the-art constructions of lossless condensers, however, we
  obtain explicit testing schemes with $O(d^{g+3} (\log d)
  \qpoly(\log n))$ and $O(d^{g+3+\beta} \poly(\log n))$ measurements,
  for arbitrary constant $\beta > 0$.
\end{abstract}

\section{Introduction}

Combinatorial group testing is a classical problem that deals with
identification of sparse Boolean vectors using disjunctive
queries. Suppose that among a large set of $n$ items it is suspected
that, for some \emph{sparsity parameter} $d \ll n$, up to $d$ items
might be ``defective''. In technical terms, defective items are known
as \emph{positives} and the rest are called \emph{negatives}. 
In a \emph{pooling strategy}, the items may be
arbitrarily grouped in pools, and a single ``measurement'' reveals
whether there is one or more positives within the chosen pool.  The
basic goal in group testing to design the pools in such a way that the
set of positives can be identified from a number of measurements
that is substantially less than $n$. 

Since its introduction in 1940's
\cite{ref:Dor43}, group testing and its variations have been
extensively studied and have found surprisingly many applications in
seemingly unrelated areas. In particular, we mention applications
in molecular biology and DNA library screening (cf.\
\cite{ref:BKBB95,ref:KKM97,ref:Mac99,ref:ND00,ref:STR03,ref:WHL06,ref:WLHD08}
and the references therein), multi-access communication
\cite{ref:Wol85}, data compression \cite{ref:HL00}, pattern matching
\cite{ref:CEPR07}, streaming algorithms \cite{ref:CM05}, software
testing \cite{ref:BG02}, compressed sensing \cite{ref:CM06}, and secure
key distribution \cite{ref:CDH07}, among others.  We refer the reader
to \cite{ref:groupTesting,ref:DH06} for an extensive review of the major
results in this area.

Formally, in classical group testing one aims to learn an unknown
Boolean vector $(x_1, \ldots, x_n) \in \zo^n$ which is known to be
$d$-sparse (that is, contains at most $d$ non-zero entries)
using a set of $m$ measurements, where each measurement is defined by a subset
of the coordinates $\mathcal{I} \subseteq [n]$ and outputs the logical
``or'' $\bigvee_{i \in \mathcal{I}} x_i$. The goal is then to design the
measurements in such a way that all $d$ sparse vectors become uniquely
identifiable using as few number of measurements as possible.

A natural generalization of classical group testing (that we call \emph{threshold testing}), 
introduced by
Damaschke \cite{ref:thresh1}, considers the case where the measurement
outcomes are determined by a \emph{threshold predicate} instead of the
logical or. Namely, this model is characterized by two integer
parameters $\ell, u$ such that $0 \geq \ell < u$ (that are considered
fixed constants), and each measurement outputs positive if the
number of positives within the corresponding pool is at least $u$. On
the other hand, if the number of positives is less than or equal to%
\footnote{
The proceedings version of this paper \cite{ref:Che10} and also
the author's Ph.D.~thesis \cite{mahdiPhD} use a slightly different
notation 
where the test returns negative if the number of positives
in the group is strictly less than $\ell$. Accordingly in those versions
the gap parameter is defined to be $u-\ell$ rather than $u-\ell-1$.
A revised notation is used in this version to make the exposition consistent
with the original paper of Damaschke \cite{ref:thresh1}.
} 
$\ell$, the
test returns negative, and otherwise the outcome can be arbitrary (that is,
either $0$ or $1$ in any arbitrary way).  In
this view, classical group testing corresponds to the special case
where $\ell = 0$ and $u = 1$.  In addition to being of theoretical interest,
the threshold model is interesting for applications, in particular
in biology, where the measurements have reduced or
unpredictable sensitivity or may depend on various factors that must
be simultaneously present in the sample to result in a positive outcome.

The difference $g := u - \ell - 1$ is known as the
\emph{gap} parameter.  As shown by Damaschke \cite{ref:thresh1}, in
threshold group testing identification of the set of positives is only
possible when the number of positives is at least $u$.  Moreover,
regardless of the number of measurements, in general the set of
positives can be only approximately identified within up to $g$ false positives and
$g$ false negatives
  (thus, unique identification can only be
guaranteed when $\ell = u - 1$). Additionally, Damaschke constructed a
scheme for identification of the positives in the threshold model. For
the gap-free case where $g=0$, the number of measurements in this
scheme is $O(d \log n)$, which is nearly optimal (within constant
factors).  However, when $g > 0$, the number of measurements becomes
$O(dn^b + d^u)$, for an arbitrary constant $b > 0$, if up to
$g+(u-1)/b$ misclassifications are allowed.

A drawback of the scheme presented by Damaschke is that the
measurements are \emph{adaptive}; i.e., the group chosen by each measurement can depend on
the outcomes of the previous ones. For numerous applications
(in particular, in molecular biology), adaptive measurements are
infeasible and must be avoided. In a \emph{non-adaptive} setting, all
measurements must be specified before their outcomes are revealed.
This makes it convenient to think of the measurements in a matrix
form. Specifically, a non-adaptive \emph{measurement matrix} is an $m
\times n$ Boolean matrix whose $i$th row is the characteristic vector
of the set of items participating in the $i$th pool, and the goal
would be to design a suitable measurement matrix.

More recently, non-adaptive threshold testing has been considered by Chen and
Fu~\cite{ref:thresh2}. They observe that a
generalization of the standard notion of disjunct matrices, the latter
being extensively used in the literature of classical group testing,
is suitable for the threshold model. Throughout this work, we refer to this
generalized notion as \emph{strongly disjunct} matrices and to the standard
notion as \emph{classical} disjunct matrices. Using strongly disjunct matrices, they
show that $O(e d^{u+1} \log(n/d))$ non-adaptive
measurements suffices to identify the set of positives (within $g$
false positives/negatives) even if up to $e$ erroneous measurements
are allowed in the model. This number of measurements almost matches
(up to constant factors) the known lower bounds on the number of rows
of strongly disjunct matrices. However, the dependence on the sparsity
parameter is $d^{u+1}$, which might be prohibitive for an interesting
range of parameters, when the thresholds are not too small (e.g.,
$\ell+1 = u=10$) and the sparsity parameter is rather large (e.g., $d =
n^{1/10}$).

In this work, we consider the non-adaptive threshold model in a
possibly noisy setting, where a number of measurement outcomes
(specified by an \emph{error parameter} $e \geq 0$) may be
incorrect. Our first observation is that, a new variation of classical
disjunct matrices (that is in general strictly weaker than strongly
disjunct matrices) suffices for the purpose of threshold group
testing. 
Using a randomness-efficient probabilistic construction (that
requires $\poly(d, \log n)$ bits of randomness), we construct
generalized disjunct matrices with $O(d^{g+2} (\log d) \log(n/d))$
rows. Thus, we bring the exponent of $d$ in the asymptotic number of
measurements from $u+1$ (that is optimal for strongly disjunct
matrices) down to $g+2$, which is \emph{independent} of the actual
choice of the thresholds and only depends on the \emph{gap} between them. We
also show 
that this tradeoff is essentially optimal for our notion of
disjunct matrices. In the gap-free case, we furthermore show that this tradeoff
is in fact the best to hope for (up to a $\log d$ term) for any threshold testing
design, and thus our notion of disjunct matrices is indeed optimal (Corollary~\ref{coro:lowerboundSimplified}).
For the positive-gap case, we show that the dependence $d^{g+2}$,
up to poly-logarithmic factors, is necessary for any
threshold testing design, and thus our notion obtains the correct
exponent (Corollary~\ref{coro:lowerboundGeneralSimplified}).

We proceed to define a new auxiliary object, namely the notion of
\emph{regular} matrices, that turns out to be the key combinatorial
object in our explicit constructions. Intuitively, given a gap $g \geq
0$, a suitable regular matrix $\cM_1$ can be used to take any
measurement matrix $\cM_2$ designed for the threshold model with lower
threshold $\ell = 0$ and higher threshold $u = g+1$ and ``lift'' it up 
to matrix that works for any
arbitrary lower threshold $\ell' > 0$ and the same gap $g$. Therefore,
for instance, in order to address the gap-free model, it would suffice to
have a non-adaptive scheme for the classical group testing model with $\ell+1 = u =
1$.  This transformation is accomplished using a simple product that
increases the height of the original matrix $\cM_2$ by a
multiplicative factor equal to the height of the regular matrix
$\cM_1$, while preserving the ``low-threshold'' distinguishing properties of the original matrix
$\cM_2$.

Next, we introduce a framework for construction of regular
matrices using \emph{strong lossless condensers} that are fundamental
objects in derandomization theory, and more generally, theoretical
computer science. We show that, by using an optimal condenser, it is
possible to construct regular matrices with only $O(d (\log d)\log n)$
rows. This almost matches the upper bound achieved by a probabilistic
construction that we also present in this work.  To this date, no explicit
construction of such optimal lossless condensers is known (though
probabilistic constructions are easy to obtain). However, using
state of the art in explicit condensers \cite{ref:CRVW02,ref:GUV07},
we will obtain two explicit constructions of regular matrices with
incomparable parameters. Namely, one with $O(d (\log d) \qpoly(\log
n))$ rows and another with $O(d^{1+\beta} \poly(\log n))$, where
$\beta > 0$ is any arbitrary constant and the exponent of the term
$\poly(\log n)$ depends on the choice of $\beta$. By combining regular
matrices with strongly disjunct ones (designed for the lowered thresholds
$\ell'=0$ and $u' = g+1$), we obtain our threshold testing schemes. The bounds
obtained by our final schemes are summarized in Table~\ref{tab:params}.
When the lower threshold $\ell$ is not too small, our explicit constructions
(rows M8 and M9 of Table~\ref{tab:params}) significantly improve what was previously known to be achievable even
using non-constructive proofs.

\begin{table}
  \caption{Summary of the parameters achieved by various constructions of threshold disjunct matrices. The 
    noise parameter $p \in [0,1)$ is arbitrary, and thresholds $\ell, u = \ell+g+1$ are fixed constants. 
    ``Exp'' and ``Rnd'' respectively indicate explicit and randomized constructions. 
    ``KS'' refers to the construction of strongly disjunct matrices based on Kautz-Singleton
    superimposed codes \cite{ref:KS64}, as described later in Section~\ref{app:strongly} (the bounds
    in rows M1-M5 are obtained by strongly disjunct matrices).
  }

  \begin{center}
    \begin{tabular}{|c|l|l|p{7cm}|}
      \hline
      & Number of rows & Tolerable errors & Remarks \\ \hline \hline
      M1 & $O(d^{u+1} \frac{\log (n/d)}{(1-p)^2})$ & $\Omega(p d \frac{\log (n/d)}{(1-p)^2})$ & Rnd: Random strongly disjunct matrices. \\
      M2 & $O((\frac{d}{1-p})^{u+1} \log n)$ & $\Omega(p d \frac{\log n}{1-p})$ & Exp: KS using codes on the GV bound. \\
      M3 & $O((\frac{d \log n}{1-p})^{u+1})$ & $\Omega(p d \frac{\log n}{1-p})$ & Exp: KS using Reed-Solomon codes. \\
      M4 & $O((\frac{d}{1-p})^{2u+1} \log n)$ & $\Omega(p d \frac{\log n}{1-p})$ & Exp: KS using Algebraic Geometric codes. \\
      M5 & $O((\frac{d \sqrt{\log n}}{1-p})^{u+3/2})$ & $\Omega(p (\frac{d \sqrt{\log n}}{1-p})^{3/2})$ & Exp: KS using Hermitian codes ($d \gg \sqrt{\log n}$). \\
      \hline \hline
      M6 & $O(d^{g+2} \frac{(\log d) \log (n/d)}{(1-p)^2})$ & $\Omega(p d \frac{\log (n/d)}{(1-p)^2})$ & Rnd: Construction~\ref{constr:prob}. \\
      M7 & $O(d^{g+3} \frac{(\log d) \log^2 n}{(1-p)^{2}})$ & $\Omega(p d^2 \frac{\log^2 n}{(1-p)^2})$ & 
      Constructions \ref{constr:reg}~and~\ref{constr:replace} combined, assuming optimal condensers and strongly disjunct matrices.\\
      M8 & $O(d^{g+3} \frac{(\log d) T_2 \log n}{(1-p)^{g+2}})$ & $\Omega(p d^2 \frac{T_2 \log n}{1-p})$ & 
      Exp: Constructions \ref{constr:reg}~and~\ref{constr:replace} combined using Theorem~\ref{thm:CRVW} and M2, where
      $T_2 = \exp(O(\log^3 \log n )) = \qpoly(\log n)$. \\
      M9 & 
      $O(d^{g+3+\beta} \frac{T_3^{\ell} \log n}{(1-p)^{g+2}})$ 
      & $\Omega(p d^{2-\beta} \frac{ \log n}{1-p} )$ & 
      Exp: Constructions \ref{constr:reg}~and~\ref{constr:replace} combined using Theorem~\ref{thm:GUVcond} and M2, where
      $\beta>0$ is any arbitrary constant and 
      $T_3 = ((\log n)(\log d))^{1+u/\beta} = \poly(\log n, \log d)$.  \\
      \hline \hline
      & $\Omega(d^{g+2} \log_d n + e d^{g+1})$ & $e$ & Lower bound (Theorem~\ref{thm:lowerBoundThresholdDisjunctGeneral}). \\
      \hline
  \end{tabular}
  \end{center}
  \label{tab:params}
\end{table}

 The rest of the paper is organized as follows. In
 Section~\ref{sec:prelim} we introduce preliminary notions and fix
 some notation. In Section~\ref{sec:disjunct} we formalize the notion
 of threshold testing designs. Moreover, we review the notion of
 strongly disjunct matrices and introduces our weaker notion of threshold disjunct
 matrices (for the
 gap-free case $g=0$), in addition to the notion of regular matrices
 and its properties. We will also prove lower bounds on the number of rows
 of such matrices. In Section~\ref{sec:constr} we obtain matching probabilistic
 upper bounds on the number of rows using the probabilistic method. Furthermore, 
 we develop our
 construction of regular matrices from lossless condensers, and
 instantiate the parameters in Section~\ref{sec:instan}.  This in
 particular leads to our explicit threshold testing schemes.  
 In Section~\ref{sec:gap} we extend all our results to the case with nonzero gap. 
 In Section~\ref{app:strongly}, we obtain explicit constructions of strongly
 disjunct matrices from error-correcting codes, by extending the classical
 technique initiated by Kautz and Singleton. Finally, in
 Section~\ref{sec:concl} we discuss the future directions.

\subsection{Preliminaries} \label{sec:prelim}

For a matrix $\cM$, we denote by $\cM[i,j]$ the entry of $\cM$ at the
$i$th row and the $j$th column.  Similarly, we denote the $i$th entry of a
vector $v$ by $v(i)$.  The \emph{support} a vector $x \in \zo^n$,
denoted by $\supp(x)$, is a subset of $[n] := \{1, \ldots, n\}$ such
that $i \in \supp(x)$ if and only if $x(i)=1$. The Hamming weight of $x$, denoted
by $\wgt(x)$ is defined as $|\supp(x)|$. The Hamming distance
between vectors $x, x' \in \zo^n$ is denoted by $\dist(x, x')$.

For an $m \times n$ Boolean matrix $\cM$ and $S \subseteq [n]$, we
denote by $\cM|_S$ the $m \times |S|$ submatrix of $\cM$ formed by
restricting $\cM$ to the columns picked by $S$.  Moreover, for a
vector $x \in \zo^n$, we use $\cM[x]_{\ell,u}$ to denote the set of
all possible outcomes of measuring $x$ in the threshold model with
lower and upper thresholds $\ell$ and $u$ and using the measurement
matrix $M$. 
Formally, for any $y \in \cM[x]_{\ell,u}$ we have $y(i) = 1$ if
$|\supp(\cM(i)) \cap \supp(x)| \geq u$, and $y(i) = 0$ if
$|\supp(\cM(i)) \cap \supp(x)| \leq \ell$, where here $\cM{(i)}$ indicates
the $i$th row of $\cM$.
In the gap-free case, the measurement outcome is uniquely defined
(since there is no ambiguity in the measurement process), and thus
the set $\cM[x]_{\ell,u}$ only contains a single element that we denote by
$\cM[x]_u$.

The \emph{min-entropy} of a distribution $\cX$ with finite support
$\Omega$ is given by \[ H_\infty(\cX) := \min_{x \in \Omega}\{-\log
\cX(x)\}, \] where $\cX(x)$ is the probability that $\cX$ assigns to
the outcome $x$ and logarithm is taken to base~$2$.  A \emph{flat} distribution is one that is uniform
on its support. For such a distribution $\cX$, we have $H_\infty(\cX) = \log(|\supp(\cX)|)$. 
The \emph{statistical distance} between two distributions
$\cX$ and $\cY$ defined on the same finite space $\Omega$ is given by
$ \frac{1}{2} \sum_{s \in \Omega} |\cX(s) - \cY(s)|, $ which is half
the $\ell_1$ distance of the two distributions when regarded as
vectors of probabilities over $\Omega$. Two distributions $\cX$ and
$\cY$ are said to be $\eps$-close if their statistical distance is at
most $\eps$.  We will use the shorthand $\U_n$ for the uniform
distribution on $\zo^n$, and $X \sim \cX$ for a random variable $X$
drawn from a distribution $\cX$. 

The main technical tool that we
use in our explicit constructions is the notion of \emph{lossless
  condensers}, defined below. 
\begin{defn}
  A function $f\colon \zo^\tn \times \zo^\tee \to \zo^\tl$ is a strong
  \emph{lossless condenser} for entropy $\tk$ and with \emph{error}
  $\eps$ (in short, $(\tk,\eps)$-condenser) if for every distribution
  $\cX$ on $\zo^\tn$ with min-entropy at least $\tk$, random variable
  $X \sim \cX$ and a \emph{seed} $Y \sim \U_\tee$, the distribution of
  $(Y, f(X, Y))$ is $\eps$-close to some distribution $(\U_\tee, \cZ)$
  with min-entropy at least $\tee+\tk$.
  A condenser is \emph{explicit} if it is polynomial-time computable.
\end{defn}

We will use the following ``almost-injectivity'' property of lossless condensers
in our proofs.

\begin{prop} \label{prop:flatmap} Let $\cX$ be a flat distribution
  with min-entropy $\log K$ over a finite sample space $\Omega$ and
  $f\colon \Omega \to \Gamma$ be a mapping to a finite set
  $\Gamma$. If $f(\cX)$ is $\eps$-close to having min-entropy $\log K$,
  then there is a set $T \subseteq \Gamma$ of size at least
  $(1-4\eps)K$ such that
  \[
  (\forall y \in T)\quad f(x)=y \land f(x') = y \Rightarrow x = x'.
  \]
\end{prop}

\begin{Proof}
  Suppose that $\cX$ is uniformly supported on a set $S \subseteq
  \Omega$ of size $K$. For each $y \in \Gamma$, define $n_y := |\{x
  \in \Omega\colon f(x) = y\}|$. Denote by $\mu$ the distribution
  $f(\cX)$ over $\Gamma$ and by $\mu'$ a distribution on $\Gamma$ with
  min-entropy $\log K$ that is $\eps$-close to $\mu$, which is guaranteed
  to exist by the assumption. Define $T := \{ y \in \Gamma\colon n_y =
  1\}$, and similarly, $T' := \{ y \in \Gamma\colon n_y \geq
  2\}$. Observe that for each $y \in \Gamma$ we have $\mu(y) = n_i/K$,
  and also $\supp(\mu) = T \cup T'$.  Thus,
  \begin{equation} \label{eqn:TTp} |T| + \sum_{y \in T'} n_y = K.
  \end{equation}
  The fact that $\mu$ and $\mu'$ are $\eps$-close implies that
  \[
  \sum_{y \in T'} | \mu(y) - \mu'(y) | \leq 2 \eps \Rightarrow \sum_{y
    \in T'} (n_y - 1) \leq 2 \eps K.
  \]
  In particular, this means that $|T'| \leq 2 \eps K$ (since by the
  choice of $T'$, for each $y \in T'$ we have $n_y \geq 2$).
  Furthermore,
  \[
  \sum_{y \in T'} (n_y - 1) \leq 2 \eps K \Rightarrow \sum_{y \in T'}
  n_y \leq 2 \eps K + |T'| \leq 4\eps K.
  \]
  This combined with \eqref{eqn:TTp} gives
  \[
  |T| = K - \sum_{y \in T'} n_y \geq (1-4\eps) K,
  \]
  as desired.
\end{Proof}

\section{Variations of disjunct matrices} \label{sec:disjunct}

The combinatorial structure used by Chen and Fu in their non-adaptive
scheme is the following generalization of the standard notion of
disjunct matrices that we refer to as \emph{strongly disjunct}
matrices throughout this work.

\begin{defn} \label{def:strongDisjunct} A matrix (with at least $d+u$
  columns) is said to be strongly $(d,e;u)$-disjunct if for every
  choice of $d+u$ columns $C_1,\ldots, C_u, C'_1, \ldots, C'_d$, all
  distinct, we have
  \[
  |\cap_{i=1}^u \supp(C_i) \setminus \cup_{i=1}^{d} \supp(C'_i) | > e.
  \]
\end{defn}
Observe that, strongly $(d,e;u)$-disjunct matrices are, in particular,
strongly $(d',e';u')$-disjunct for any $d' \leq d$, $e' \leq e$, and $u' \leq
u$.  Moreover, \emph{classical} $(d,e)$-disjunct matrices that
are extensively used in group testing literature (see
\cite[Ch.~7]{ref:groupTesting}) are equivalent to strongly
$(d,e;1)$-disjunct matrices.


To make the main ideas more transparent, until Section~\ref{sec:gap}
we will focus on the gap-free case where $\ell = u-1$. The extension to
nonzero gaps is straightforward and will be discussed in
Section~\ref{sec:gap}. Moreover, often we will implicitly assume that
the Hamming weight of the Boolean vector that is to be identified is
at least $u$ (since otherwise, we know from the work of 
Damaschke \cite{ref:thresh1} that confusions cannot be
avoided), and will take the thresholds $\ell, u$ to be fixed constants.

The notion of strongly disjunct matrices, in its general form, has
been studied in the literature under different names and equivalent
formulations, e.g., superimposed $(u,d)$-designs/codes,
superimposed distance codes, and $(u,d)$
cover-free families (see
\cite{ref:CDH07,ref:CFH08,ref:DRR89,ref:DVMT02,ref:KL04,ref:SW00,ref:SW04} and
the references therein).  An important motivation for the study of
this notion is the following \emph{hidden hypergraph-learning problem}
(cf.\ \cite[Ch.~12]{ref:groupTesting}), itself being motivated 
by the so-called \emph{complex model} in computational biology \cite{ref:CDH07}: 
Suppose that $G$ is a $u$-hypergraph; that is, a hypergraph where
each edge is a set of $u$ vertices.
on a vertex set $V$ of size $n$, and denote by $\mathcal{V}(G)$ the
set of vertices induced by the hyper-edge set of $G$; i.e., $v \in
\mathcal{V}(G)$ if and only if $G$ has a hyper-edge incident to $v$. Then
assuming that $|\mathcal{V}(G)| \leq d$ for a \emph{sparsity
  parameter} $d$, the aim is to learn $G$ using as few (non-adaptive)
queries of the following type as possible: Each query specifies a set
$Q \subseteq V$, and its corresponding answer is a Boolean value which
is $1$ if and only if $G$ has a hyperedge contained in $Q$.  
It is known that
\cite{ref:GHTWZ06,ref:CDH07}, in the hypergraph-learning problem, any
suitable grouping strategy defines a strongly disjunct matrix (whose
rows are characteristic vectors of individual queries $Q$), and
conversely, any strongly disjunct matrix can be used as the incidence
matrix of the set of queries. The parameter $e$ determines ``noise
tolerance'' of the measurement scheme.  Namely, a strongly
$(d,e;u)$-disjunct matrix can uniquely distinguish between $d$-sparse
hypergraphs even in presence of up to $\lfloor e/2 \rfloor$
erroneous query outcomes.

For gap-free threshold group testing, the successful strategy needed for distinguishing
between $d$-sparse Boolean vectors can trivially be captured by the following
definition.

\begin{defn} \label{def:thresholdDesign}
Let $n \geq d \geq u > 0$ and $e \geq 0$ be integer parameters.
A Boolean matrix $M$ with $n$ columns is said to be a \emph{$(d, e; u)$-threshold design}
if for every $d$-sparse $x, x' \in \zo^n$ of Hamming weight $u$ or more 
such that $x \neq x'$, we have $\dist(M[x]_u, M[x']_u) > e$.
\end{defn}

The key observation made by Chen and Fu~\cite{ref:thresh2} is that
threshold group testing corresponds to the special case of the
hypergraph learning problem where the hidden graph $G$ is known to be
a $u$-clique\footnote{As standard in graph theory, a $u$-clique on the vertex set $V$ is a
  $u$-hypergraph $(V,E)$ such that, for some $V' \subseteq V$, $E$ is
  the set of all subsets of $V'$ of size $u$.}. In this case, the
unknown Boolean vector in the corresponding threshold testing problem
would be the characteristic vector of $\mathcal{V}(G)$.  It follows that
strongly disjunct matrices are threshold designs as defined in Definition~\ref{def:thresholdDesign}
Specifically,

\begin{thm}\cite{ref:thresh2} \label{thm:ChenFu}
Let $M$ be a Boolean matrix with $n$ columns that is strongly
$(d,e;u)$-disjunct. Then, $M$ is a $(d,e;u)$-threshold design. \qed
\end{thm}

Nonconstructively, a probabilistic argument akin to the standard
argument for the case of classical disjunct matrices (see
\cite[Ch.~7]{ref:groupTesting}) can be used to show that strongly
$(d,e;u)$-disjunct matrices exist with $m = O(d^{u+1} (\log
(n/d))/(1-p)^2)$ rows and error tolerance $e = \Omega(p d \log (n/d)/(1-p)^2)$,
for any noise parameter $p \in [0,1)$. On the negative side, however,
several concrete lower bounds are known for the number of rows of such
matrices \cite{ref:SW00,ref:DVMT02,ref:SW04}.  In asymptotic terms,
these results show that one must have $m = \Omega(d^{u+1} \log_d n + e
d^u)$, and thus, the probabilistic upper bound is essentially
optimal. 

For the underlying strongly disjunct matrix, Chen and Fu~\cite{ref:thresh2} use
a greedy construction \cite{ref:CFH08} that achieves, for any $e \geq 0$, $O((e+1) d^{u+1}
\log(n/d))$ rows, but may take exponential time in the size of the
resulting matrix. Nevertheless, as observed by several researchers
\cite{ref:DVMT02,ref:KL04,ref:GHTWZ06,ref:CDH07}, a classical explicit
construction of combinatorial designs due to Kautz and
Singleton~\cite{ref:KS64} can be extended to construct strongly disjunct
matrices. This concatenation-based construction transforms any
error-correcting code having large distance into a disjunct matrix.
While the original construction uses Reed-Solomon codes and achieves
nice bounds, it is possible to use other families of codes. In
particular, as recently shown by Porat and Rothschild
\cite{ref:PR08}, codes on the Gilbert-Varshamov bound
(cf. \cite{ref:MS}) result in nearly optimal disjunct
matrices. Moreover, for a suitable range of parameters, they give a
\emph{deterministic} construction of such codes that runs in
polynomial time in the size of the resulting disjunct matrix (albeit
exponential in the dimension of the code\footnote{In this regard, this
  construction of disjunct matrices can be considered \emph{weakly
    explicit} in that, contrary to fully explicit constructions, it is
  not clear if each individual entry of the matrix can be computed in
  time $\poly(d, \log n)$.  }). We will elaborate on details of this
class of constructions in Section~\ref{app:strongly}, and will
additionally consider a family of algebraic-geometric codes and
Hermitian codes which give incomparable bounds, as summarized
in Table~\ref{tab:params} (rows M2--M5).

\subsection{Threshold disjunct and regular matrices}

Even though, as discussed above, the general notion of strongly
$(d,e;u)$-disjunct matrices is sufficient for threshold group testing
with upper threshold $u$, in this section we show that a new,
weaker, notion of disjunct matrices defined below (which, as we show later, turns
out to be \emph{strictly} weaker when $u>1$), would also suffice.
We also define an auxiliary notion of \emph{regular} matrices.

\begin{defn} \label{def:regularMatrix} A Boolean matrix $\cM$ with $n$
  columns is called $(d,e;u)$-regular if for every subset of columns
  $S \subseteq [n]$ (called the \emph{critical set}) and every $Z
  \subseteq [n]$ (called the \emph{zero set}) such that $u \leq |S|
  \leq d$, $|Z| \leq |S|$, $S \cap Z = \emptyset$, there are more than
  $e$ rows of $\cM$ at which $\cM|_S$ has weight exactly $u$ and (at the same rows)
  $\cM|_Z$ has weight zero. Any such row is said to
  \emph{$u$-satisfy} $S$ and $Z$. If, in addition, for every
  \emph{distinguished column} $i \in S$, more than $e$ rows of $\cM$
  both $u$-satisfy $S$ and $Z$ and have a $1$ at the $i$th column, the
  matrix is called threshold $(d,e;u)$-disjunct (and the corresponding ``good''
  rows are said to $u$-satisfy $i$, $S$, and $Z$).
\end{defn}


To distinguish between the above variant of disjunct matrices and
strongly disjunct matrices or classical disjunct matrices, we will refer to our variant as 
\emph{threshold disjunct} matrices throughout the paper.

It is easy to verify that (assuming $2d \leq n$) the classical notion
of $(2d-1,e)$-disjunct matrices is equivalent to strongly
$(2d-1,e;1)$-disjunct and threshold $(d,e;1)$-disjunct.  Moreover, any
threshold $(d,e;u)$-disjunct matrix is $(d,e;u)$-regular, $(d-1,e;u-1)$-regular,
and classical $(d,e)$-disjunct (but the reverse implications do not in general
hold). Therefore, the known lower bound of $m = \Omega(d^{2}
\log_d n + ed)$ that applies for $(d,e)$-disjunct matrices holds for
threshold $(d,e;u)$-disjunct matrices as well (see Theorem~\ref{thm:lowerBoundThrDisjunct}).  
Below we show that our 
notion of disjunct matrices suffices for threshold designs.

\begin{lem} \label{lem:disjunct} Let $\cM$ be an $m \times n$ Boolean
  matrix that is threshold $(d,e;u)$-disjunct. Then for every distinct
  $d$-sparse vectors $x, x' \in \zo^n$ such that
    $\supp(x) \nsubseteq \supp(x')$,
  $\wgt(x) \geq |\supp(x') \setminus \supp(x)|$ and $\wgt(x) \geq u$,
  we have
  \begin{equation} \label{eqn:distin} |\supp(\cM[x]_u) \setminus
    \supp(\cM[x']_u)| > e.
  \end{equation}
  Moreover, $\cM$ is a $(d,e;u)$-threshold design.
  Conversely, if $\cM$ satisfies \eqref{eqn:distin} for every choice
  of $x$ and $x'$ as above, it must be threshold $(\lfloor d/2
  \rfloor,e;u)$-disjunct.
\end{lem}

\begin{Proof}
  First, suppose that $\cM$ is threshold $(d,e;u)$-disjunct, and let $y :=
  \cM[x]_u$ and $y' := \cM[x']_u$.  Take any $i \in \supp(x) \setminus
  \supp(x')$, and let $S := \supp(x)$ and $Z := \supp(x') \setminus
  \supp(x)$.  Note that $|S| \leq d$ and by assumption, we have $|Z|
  \leq |S|$.
  Now, Definition~\ref{def:regularMatrix} implies that there is a set
  $E$ of more than $e$ rows of $M$ that $u$-satisfy $i$ as the
  distinguished column, $S$ as the critical set and $Z$ as the zero
  set. Thus for every $j \in E$, the $j$th row of $\cM$ restricted to
  the columns chosen by $\supp(x)$ must have weight exactly $u$, while
  its weight on $\supp(x')$ is less than $u$. Therefore, $y(j) = 1$
  and $y'(j) = 0$ for more than $e$ choices of $j$.

  The claim that $\cM$ is a $(d,e;u)$-threshold design follows from
  the above argument combined with the observation that at least 
  one of the two possible orderings of any two distinct
    $d$-sparse vectors, at least one having weight $u$ or more,
    satisfies the conditions required by the lemma.

  For the converse, consider any choice of a distinguished column $i
  \in [n]$, a critical set $S \subseteq [n]$ containing $i$ (such that
  $|S| \geq u$), and a zero set $Z \subseteq [n]$ where $|Z| \leq
  |S|$.  Define $d$-sparse Boolean vectors $x, x' \in \zo^n$ so that
  $\supp(x) := S$ and $\supp(x') := Z \cup (S \setminus \{ i\})$.  Let
  $y := \cM[x]_u$ and $y' := \cM[x']_u$ and $E := \supp(y) \setminus
  \supp(y')$. By assumption we know that $|E| > e$. Take any $j \in
  E$. Since $y(j) = 1$ and $y'(j) = 0$, we get that the $j$th row of
  $\cM$ restricted to the columns picked by $Z \cup (S \setminus \{i\})$
  must have weight at most $u-1$, whereas it must have weight at least
  $u$ when restricted to $S$.  As the sets $\{i\}, S \setminus \{i\}$,
  and $Z$ are disjoint, this can hold only if $\cM[j, i] = 1$, and
  moreover, the $j$th row of $\cM$ restricted to the columns picked by
  $S$ (resp., $Z$) has weight exactly $u$ (resp., zero).  Hence, this
  row (as well as all the rows of $\cM$ picked by $E$) must $u$-satisfy
  $i, S$, and $Z$, confirming that $\cM$ is threshold $(\lfloor d/2
  \rfloor,e;u)$-disjunct.
\end{Proof}

We point out that Lemma~\ref{lem:disjunct} proves a matching converse,
suggesting that the notion of threshold disjunct matrices might be ``close'' 
to a characterization of threshold designs (Definition~\ref{def:thresholdDesign}),
up to a constant factor in the sparsity parameter. However, this does not
imply a precise characterization since the assumptions of Lemma~\ref{lem:disjunct}
consider a particular ordering on the sparse vectors $x$ and $x'$, which
must be consistent with the ordering in \eqref{eqn:distin}. 
However, as we show in Section~\ref{sec:lower}, threshold designs (Definition~\ref{def:thresholdDesign})
and threshold disjunct matrices (Definition~\ref{def:regularMatrix}) satisfy the
same asymptotic lower bounds on the number of rows, which nearly matches the
upper bounds that we prove by probabilistic arguments (Lemma~\ref{lem:probDisjunct}),
assuming that the threshold parameter is an absolute constant.
Thus, quantitatively, our notion of threshold disjunct matrices 
essentially provides an optimal way of constructing threshold group testing designs.

\begin{constr}[b] 
  \caption{Direct product of measurement matrices.}

  \begin{itemize}
  \item {\it Given: } Boolean matrices $\cM_1$ and $\cM_2$ that are
    $m_1 \times n$ and $m_2 \times n$, respectively.

  \item {\it Output: } An $m \times n$ Boolean matrix $\cM_1 \rep
    \cM_2$, where $m := m_1 m_2$.

  \item {\it Construction: } Let the rows of $\cM := \cM_1 \rep \cM_2$
    be indexed by the set $[m_1] \times [m_2]$. Then the row
    corresponding to $(i,j)$ is defined as the bit-wise or of the
    $i$th row of $\cM_1$ and the $j$th row of $\cM_2$.
  \end{itemize}
  \label{constr:replace}
\end{constr}

\subsection{Direct product of matrices}

We will use regular matrices as intermediate building blocks in our
constructions of disjunct matrices to follow. The connection with
disjunct matrices is made apparent through a direct product of matrices defined
in Construction~\ref{constr:replace}.
Intuitively, using this product, regular matrices can be used to transform any measurement
matrix suitable for the standard group testing model to one with
comparable properties in the threshold model. The following lemma
formalizes the idea.

\begin{lem} \label{lem:rep} Let $\cM_1$ and $\cM_2$ be Boolean
  matrices with $n$ columns, such that $\cM_1$ is
  $(d-1,e_1;u-1)$-regular. Let $\cM := \cM_1 \rep \cM_2$, and suppose
  that for $d$-sparse Boolean vectors $x, x' \in \zo^n$ such that
  $\wgt(x) \geq \wgt(x')$, we have
  \[
  | \supp(\cM_2[x]_1) \setminus \supp(\cM_2[x']_1)| \geq e_2.
  \]
  Then, $ |\supp(\cM[x]_u) \setminus \supp(\cM[x']_u)| \geq (e_1+1)
  e_2.  $
\end{lem}

  \begin{Proof}
  First we consider the case where $u > 1$.  Let $y := \cM_2[x]_1 \in
  \zo^{m_2}$, $y' := \cM_2[x']_1 \in \zo^{m_2}$, where $m_2$ is the
  number of rows of $\cM_2$, and let $E := \supp(y) \setminus
  \supp(y')$.  By assumption, $|E| \geq e_2$. Fix any $i \in E$ so
  that $y(i) = 1$ and $y'(i) = 0$.
  Therefore, the $i$th row of $\cM_2$ must have all zeros at positions
  corresponding to $\supp(x')$ and there is a $j \in \supp(x)
  \setminus \supp(x')$ such that $\cM_2[i, j] = 1$.  Define $S :=
  \supp(x) \setminus \{j\}$, $Z := \supp(x') \setminus \supp(x)$, $z
  := \cM[x]_u$ and $z' := \cM[x']_u$.

  As $\wgt(x) \geq \wgt(x')$, we know that $|Z| \leq |S|+1$.  The
  extreme case $|Z| = |S|+1$ only happens when $x$ and $x'$ have
  disjoint supports, in which case one can remove an arbitrary element
  of $Z$ to ensure that $|Z| \leq |S|$ and the following argument
  (considering the assumption $u>1$) still goes through.  
  
  By the
  definition of regularity, there is a set $E_1$ consisting of at
  least $e_1+1$ rows of $\cM_1$ that $(u-1)$-satisfy the critical set
  $S$ and the zero set $Z$. Pick any $k \in E_1$, and observe that $z$
  must have a $1$ at position $(k,i)$. This is because the row of
  $\cM$ indexed by $(k,i)$ has a $1$ at the $j$th position (since the
  $k$th row of $\cM_2$ does), and at least $u-1$ more $1$'s at
  positions corresponding to $\supp(x) \setminus \{j\}$ (due to
  regularity of $\cM_1$).  On the other hand, note that the $k$th row
  of $\cM_1$ has at most $u-1$ ones at positions corresponding to
  $\supp(x')$ (because $\supp(x') \subseteq S \cup Z$), and the $i$th
  row of $\cM_2$ has all zeros at those positions (because $y'(i)=0$).
  This means that the row of $\cM$ indexed by $(k,i)$ (which is the
  bit-wise or of the $k$th row of $\cM_1$ and the $i$th row of
  $\cM_2$) must have less than $u$ ones at positions corresponding to
  $\supp(x')$, and thus, $z'$ must be $0$ at position $(k,i)$.
  Therefore, $z$ and $z'$ differ at position $(k,i)$.

  Since there are at least $e_2$ choices for $i$, and for each choice
  of $i$, at least $e_1+1$ choices for $k$, we conclude that in at
  least $(e_1+1) e_2$ positions, $z$ has a one while $z'$ has a zero.

  The argument for $u=1$ is similar, in which case it suffices to take
  $S := \supp(x)$ and $Z := \supp(x') \setminus \supp(x)$.
  \end{Proof}

As a corollary it follows that, when $\cM_1$ is a
$(d-1,e_1;u-1)$-regular and $\cM_2$ is a classical $(d,e_2)$-disjunct matrix,
the product $\cM := \cM_1 \rep \cM_2$ will distinguish between any two
distinct $d$-sparse vectors (of weight at least $u$) in at least
$(e_1+1)(e_2+1)$ positions of the measurement outcomes.  This combined
with Lemma~\ref{lem:disjunct} would imply that $\cM$ is, in particular, threshold $(\lfloor d/2
\rfloor, (e_1+1) (e_2+1) -1; u)$-disjunct.  However, using a direct
argument similar to the above lemma it is possible to obtain a
slightly better result, given by Lemma~\ref{lem:repdisj}.

\begin{lem} \label{lem:repdisj} Suppose that $\cM_1$ is a
  $(d,e_1;u-1)$-regular and $\cM_2$ is a classical $(2d,e_2)$-disjunct matrix.
  Then $\cM_1 \rep \cM_2$ is a threshold $(d, (e_1+1) (e_2+1) -1; u)$-disjunct
  matrix. \qed
\end{lem}

As a particular example of where Lemma~\ref{lem:rep} can be used, 
we remark that the measurement matrices 
constructed in \cite{ref:Che09} that are not
necessarily disjunct but allow approximation of sparse vectors in 
highly noisy settings of the standard group testing model (as well
as those used in adaptive two-stage schemes; cf.\ \cite{ref:CD08} and
the references therein), can be
combined with regular matrices to offer the same qualities in the
threshold model. In the same way, numerous existing results in group testing
can be ported to the threshold model by using Lemma~\ref{lem:rep}.

\subsection{Lower bounds} \label{sec:lower}

In this section, we show that the known asymptotic lower bounds on the number
of rows of classical disjunct matrices apply to threshold designs (Definition~\ref{def:thresholdDesign})
and our notion of threshold disjunct matrices (\ref{def:regularMatrix}) as well.
It is immediate from the definitions that, assuming $2d \leq n$, a
threshold $(d, e; u)$-disjunct matrix is in particular a classical
$(d, e)$-disjunct matrix. Thus the latter lower bound is straightforward.

\begin{thm} \label{thm:lowerBoundThrDisjunct}
For every integer $d > 0$ there is an $n_0 > 0$ such that the following holds.
For any $n \geq n_0$, let $M$ be an $m \times n$ threshold $(d, e; u)$-disjunct
matrix. Then,
\[ m = \Omega( d^2 \log_d n + de ). \]
\end{thm}

\begin{proof}
Immediate from the known bounds on the number of rows
of classical disjunct matrices (e.g., Theorem~2.19 of \cite{ref:SW04}).
\end{proof}

Now, in order to show that any $(d, e; u)$-threshold design must satisfy
essentially the same lower bound as in Theorem~\ref{thm:lowerBoundThrDisjunct},
we first observe the following combinatorial property of such matrices.

\begin{lem} \label{lem:necessity}
Let $M$ be a $(d+1,e;\ell+1)$-threshold design.
Then it satisfies the following property:

``For every $S \subseteq [n]$ such that $|S| = d$ and every $i \in [n] \setminus S$,
there are more than $e$ rows of $M$ at which the $i$th column of $M$ contains a $1$ and
moreover in those rows, $M|_S$ has weight exactly $\ell$.''
\end{lem}

\begin{proof}
This is a special case of Lemma~\ref{lem:necessityGeneral} that will be proved later (it
suffices to set $u=\ell+1$ and $g=1$ in Lemma~\ref{lem:necessityGeneral}).
\end{proof}

\begin{thm} \label{thm:lowerboundGapfree}
For every integer $d > 0$ there is an $n_0 > 0$ such that the following holds.
For any $n \geq n_0$, let $M$ be an $m \times n$ Boolean matrix that satisfies the
property quoted in Lemma~\ref{lem:necessity}. Then,
\[ m = \Omega\Big( \big(\frac{d}{\ell+1}\big)^2 \log_d n + \frac{de}{(\ell+1)^2} \Big). \]
\end{thm}

\begin{proof}
We reduce the matrix to a classical disjunct matrix, and use the existing
lower bounds. Let $d' := \lfloor d/(\ell+1) \rfloor$ and $e' := e/(\ell+1)$. 
We define the following notation: For a set $S \subseteq [n]$
and $i \in [n] \setminus S$, a vector $v \in \zo^n$ is said to \emph{satisfy} $(i,S)$ 
if $v(i) = 1$ and $v(j) = 0$ for all $j \in S$.

For each $i \in [n]$,
we create a set $T(i) \subseteq [n]$ according to the following greedy
algorithm:

\begin{enumerate}
\item Initialize $T(i)$ with the empty set.

\item Let $S \subseteq [n] \setminus (T(i) \cup \{i\})$ be any set of size at most $d'$ such that
the number of rows of $M$ that satisfy $(i,S)$ is at most $e'$. If there
is no such $S$, terminate.

\item Set $T(i) := T(i) \cup S$, and go to step $2$.
\end{enumerate}

First, we argue that the above algorithm always terminates after looping at
most $\ell$ times. Suppose for the sake of contradiction that 
the algorithm loops more and let $S_1, \ldots, S_{\ell+1}$
be the disjoint sets $S$ obtained in the first $\ell+1$ iterations of the loop.
Let $M'$ be the matrix obtained from $M$ by removing all the rows 
where the $i$th column has a zero, and define $T' := S_1 \cup \ldots \cup S_{\ell+1}$.

By the way the algorithm chooses the sets $S_j$, we know for each 
$S_j$ that all but at most $e'$ rows of $M'|_{S_j}$
have nonzero weights. Therefore, all but at most $e'(\ell+1) = e$
rows of $M'|_{T}$ have weights at least $\ell+1$ (i.e., at least
one nonzero entry for the range of each $S_j$). 

On the other hand, since $|S_j| \leq d'$
for all $j$, we have $|T'| \leq d'(\ell+1) \leq d$. So, 
the property of Lemma~\ref{lem:necessity} implies that
there are more than $e$ rows of $M'$ where $M'|_{T'}$ has
weight exactly $\ell$.  This is a contradiction.
Therefore, we conclude, for every $i$, that $|T(i)| \leq \ell d' < d$.

Now, define an undirected graph $G=(V,E)$ where $V := [n]$ and 
$\{i,j\} \in E$ iff either $j \in T(i)$ or $i \in T(j)$.
We know from the upper bound on the size of every $T(i)$ that
the maximum degree of this graph is less than $2d$.
Therefore, the graph has an independent set $V' \subseteq V$ of size at least
$n/(2d)$. Let $M'' := M|_{V'}$, with columns indexed by the elements of $V'$.

Now, consider any $i \in V'$ and any set $S \subseteq V' \setminus i$ where
$|S| = d'$. Since $V'$ is an independent set of $G$, we know that
$T(i) \cap V' = \emptyset$. Since the greedy algorithm, given input $i$, has terminated
at step $2$, we know that there are more than $e'$ rows of $M''$ that
satisfy $(i,S)$ (otherwise the algorithm would add $S$ to $T(i)$ and loop another time). 
Since this holds for every choice of $(i,S)$, we conclude that the matrix
$M''$ must be a classical $(d',e')$-disjunct matrix. 

Let $n'$ be the number of columns
of $M''$, so we know that $n' \geq n/(2d)$.
Now it suffices to apply the known asymptotic lower bounds for
the number of rows of classical disjunct matrices
\cite{ref:DR83,ref:Rus94,ref:SW04} on $M''$. In particular, 
Theorem~2.19 of \cite{ref:SW04} implies that, 
for some absolute constant $c > 0$, and whenever $n$ is sufficiently large
for the given parameter $d$, 
\begin{eqnarray*}
m &\geq& 0.7 c \frac{(d'+1)^2}{\log(d'+1)} \log n' + 0.5 c (d'+1) e' \\
&=& \Omega\big( \frac{d^2 (\log n - \log d - 1)}{(\ell+1)^2 \log d} + \frac{de}{(\ell+1)^2}  \big),
\end{eqnarray*}
which implies the claimed bound assuming $n$ is large enough.
\end{proof}

\begin{coro} \label{coro:lowerboundSimplified}
For every integer $d > 0$ there is an $n_0 > 0$ such that the following holds.
For any $n \geq n_0$, let $M$ be an $m \times n$ Boolean matrix that is a
\emph{$(d, e; u)$-threshold design}, for some constant $u > 0$.
Then,
\[
m = \Omega_u(d^2 \log_d n + de). 
\]
\end{coro}

\begin{proof}
Immediate from Lemma~\ref{lem:necessity} and Theorem~\ref{thm:lowerboundGapfree}.
\end{proof}

\section{Constructions} \label{sec:constr}

In this section, we obtain several construction of regular and disjunct
matrices. Our first construction, described in Construction~\ref{constr:prob}, is a randomness-efficient 
probabilistic construction that can be
analyzed using standard techniques from the probabilistic method. The bounds
obtained by this construction are given in Lemma~\ref{lem:probDisjunct} below.
The amount of random bits required by this
construction is polynomially bounded in $d$ and $\log n$, which is
significantly smaller than it would be had we picked the entries of
$\cM$ fully independently.

\begin{constr}[b] 
  \caption{Probabilistic construction of regular and disjunct
    matrices.}

  \begin{itemize}
  \item {\it Given: } Integer parameters $n, m', d, u$.

  \item {\it Output: } An $m \times n$ Boolean
    matrix $\cM$, where $m := m' \lceil \log(d/u) \rceil$.

  \item {\it Construction: } Let $r := \lceil \log(d/u) \rceil$. Index
    the rows of $\cM$ by $[r] \times [m']$.
%
    Sample the $(i, j)$th row of $\cM$ independently from a
    $(u+1)$-wise independent distribution on $n$ bit vectors, where
    each individual bit has probability $1/(2^{i+2} u)$ of being $1$.
  \end{itemize}
  \label{constr:prob}
\end{constr}

\begin{lem} \label{lem:probDisjunct} For every $p \in [0,1)$ and
  integer parameter $u > 0$, Construction~\ref{constr:prob} with
  $m' = O_{u}(d \log (n/d)/(1-p)^2)$ (resp., $m' = O_{u}(d^2 \log
  (n/d)/(1-p)^2)$) outputs a $(d, \Omega_{u}(pm');u)$-regular (resp.,
  threshold $(d,\Omega_{u}(pm'/d);u)$-disjunct) matrix with probability $1-o(1)$.
\end{lem}

 \begin{Proof}
  We show the claim for regular matrices, the proof for disjunct
  matrices is similar.  Consider any particular choice of a critical
  set $S \subseteq [n]$ and a zero set $Z \subseteq [n]$ such that $u
  \leq |S| \leq d$ and $|Z| \leq |S|$. Choose an integer $i$ so that
  $2^{i-1} u \leq |S| \leq 2^i u$, and take any $j \in [m']$. Denote
  the $(i,j)$th row of $\cM$ by the random variable $\rv{w} \in \zo^n$, and
  by $q$ the ``success'' probability that $\rv{w}|_S$ has weight
  exactly $u$ and $\rv{w}|_Z$ is all zeros.  For an integer $r >
  0$, we will use the shorthand $1^r$ (resp., $0^r$) for the
  all-ones (resp., all-zeros) vector of length $r$.  We have

\begin{eqnarray}
  q &=& \sum_{\substack{R \subseteq [S] \\ |R| = u}} \Pr[(\rv{w}|_R) = 1^u \land (\rv{w}|_{Z \cup (S \setminus R)}) = 0^{|S|+|Z|-u}] \nonumber \\
  &=& \sum_{R} \Pr[(\rv{w}|_R) = 1^u] \cdot \Pr[(\rv{w}|_{Z \cup (S \setminus R)}) = 0^{|S|+|Z|-u} \mid (\rv{w}|_R) = 1^u] \nonumber \\
  &\stackrel{\mathrm{(a)}}{=}& \sum_{R} (1/(2^{i+2} u))^u \cdot (1-\Pr[(\rv{w}|_{Z \cup (S \setminus R)}) \neq 0^{|S|+|Z|-u} \mid (\rv{w}|_R) = 1^u]) \nonumber \\
  &\stackrel{\mathrm{(b)}}{\geq}& \sum_{R} (1/(2^{i+2} u))^u \cdot (1-(|S|+|Z|-u)/(2^{i+2} u)) \nonumber \\
  &\geq& \frac{1}{2} \binom{|S|}{u} (1/(2^{i+2} u))^u \geq \frac{1}{2} \left(\frac{|S|}{u}\right)^u \cdot (1/(2^{i+2} u))^u \geq \frac{1}{2^{3u+1} \cdot u^u} =: c, \label{eqn:succesProb}
\end{eqnarray}
where $\mathrm{(a)}$ and $\mathrm{(b)}$ use the fact that the entries of $\rv{w}$
are $(u+1)$-wise independent, and $\mathrm{(b)}$ uses an additional
union bound. Here the lower bound $c > 0$ is a constant that only
depends on $u$.  Now, let $e := m'p q$. using Chernoff bounds, and
independence of the rows, the probability that there are at most $e$
rows (among $(i,1), \ldots, (i, m')$) whose restrictions to $S$ and $Z$
have weights $u$ and $0$, respectively, becomes upper bounded by
\[
\exp( -(m'q-e)^2/(2m'q) ) = \exp( -(1-p)^2 m' q/2 ) \leq \exp( -(1-p)^2 m'
c/2 ).
\]
Now take a union bound on all the choices of $S$ and $Z$ to conclude
that the probability that the resulting matrix is not $(d,e;
u)$-regular is at most
\[
\left(\sum_{s=u}^{d} \binom{n}{s} \sum_{z=0}^{s} \binom{n-s}{z}\right)
\exp( -(1-p)^2 m' c/2 ),
\]
which can be made $o(1)$ by choosing $m' = O_{u}(d \log(n/d)/(1-p)^2)$.

The proof of the claim for disjunct matrices follows along the same lines, except
that we additionally need the vector $\rv{w}$ to be $1$ at the position corresponding
to the distinguished column $i$. Under this additional requirement, the
lower bound on $q$ would become $\Omega_u(1/d)$, and this only increases the number
of rows by a factor $O_u(d)$.
\end{Proof}

A significant part of this work is a construction of regular matrices using strong 
lossless condensers. Details of the construction are described in
Construction~\ref{constr:reg} that assumes a family of lossless condensers with
different entropy requirements\footnote{We have assumed
      that all the functions in the family have the same seed length
      $t$. If this is not the case, one can trivially set $t$ to be
      the largest seed length in the family.}, and in turn, uses Construction~\ref{constr:main}
as a building block.
The theorem below analyzes the obtained parameters
without specifying any particular choice for the underlying family of
condensers. 

\begin{constr} 
  \caption{A building block for construction of regular matrices.}

  \begin{itemize}
  \item {\it Given: } A strong lossless $(\tk, \eps)$-condenser
    $f\colon \zo^\tn \times \zo^\tee \to \zo^\tl$, integer parameter 
    $u \geq 1$ and real parameter $p \in
    [0,1)$ such that $\eps < (1-p)/32$, 

  \item {\it Output: } An $m \times n$ Boolean matrix $\cM$, where $n
    := 2^\tn$ and $ m = 2^{\tee+\tk} O_u(2^{u(\tl-\tk)}) $.

  \item {\it Construction: }
%
    Let $G_1=(\zo^\tl, \zo^\tk, E_1)$ be any bipartite bi-regular
    graph with left vertex set $\zo^\tl$, right vertex set $\zo^\tk$,
    edge set $E_1$, left degree $d_\ell := 8u$, and right degree $d_r := 8u
    2^{\tl-\tk}$.  Replace each right vertex $v$ of $G_1$ with
    $\binom{d_r}{u}$ vertices, one for each subset of size $u$ of the
    vertices on the neighborhood of $v$, and connect them to the vertices in the
    corresponding subsets. Denote the resulting graph by $G_2 =
    (\zo^\tl, V_2, E_2)$, where $|V_2| = 2^\tk \binom{d_r}{u}$ and
    $E_2$ is the edge set of the graph.
    Define the bipartite graph $G_3=(\zo^n, V_3, E_3)$, where $V_3 :=
    \zo^t \times V_2$ is the set of right vertices, 
    as follows: Each left vertex $x \in \zo^n$ is
    connected to $(y, \Gamma_2(f(x,y))$, for each $y \in \zo^t$, where
    $\Gamma_2(\cdot)$ denotes the neighborhood function of $G_2$
    (i.e., $\Gamma_2(v)$ denotes the set of vertices adjacent to $v$
    in $G_2$).  The output matrix $\cM$ is the bipartite adjacency
    matrix of $G_3$ with columns indexed by the left vertices of
    row indexed by the right vertices of the graph.
  \end{itemize}
  \label{constr:main}
\end{constr}

\begin{constr} 
  \caption{Regular matrices from strong lossless condensers.}

  \begin{itemize}
  \item {\it Given: } Integer parameters $d \geq u \geq 1$, real
    parameter $p \in [0,1)$, and a family $f_0, \ldots, f_r$ of
    strong lossless condensers, where $r := \lceil \log(d/u') \rceil$ and $u'$ is the
    smallest power of two such that $u' \geq u$.  Each $f_i\colon
    \zo^\tn \times \zo^\tee \to \zo^{\tl(i)}$ is assumed to be a
    strong lossless $(\tk(i), \eps)$-condenser, where $\tk(i) := \log
    u'+i+1$ and $\eps < (1-p)/32$.

  \item {\it Output: } An $m \times n$ Boolean matrix $\cM$, where $n
    := 2^\tn$ and $
    m = 2^{\tee} d \sum_{i=0}^r O_u(2^{u (\tl(i)-\tk(i))}) $.

  \item {\it Construction: } For each $i \in \{0, \ldots, r\}$, denote
    by $\cM_i$ the output matrix of Construction~\ref{constr:main}
    when instantiated with $f_i$ as the underlying condenser, and by
    $m_i$ its number of rows.  Define $r_i := 2^{r-i}$ and let
    $\cM'_i$ denote the matrix obtained from $\cM_i$ by repeating each
    row $r_i$ times.  Construct the output matrix $\cM$ by stacking
    $\cM'_0, \ldots, \cM'_r$ on top of one another.
  \end{itemize}
  \label{constr:reg}
\end{constr}

\begin{thm} \label{thm:regular} The $m \times n$ matrix $\cM$ output
  by Construction~\ref{constr:reg} is $(d,p \gamma 2^t;u)$-regular,
  where $\gamma = \max\{1, \Omega_u(d \cdot \min\{2^{\tk(i)-\tl(i)}
  \colon i=0,\ldots,r \}) \}$.
\end{thm}

\begin{Proof}
As a first step, we verify the upper bound on the number of
measurements $m$. Each matrix $\cM_i$ has $m_i = 2^{t+\tk(i)}
O_u(2^{u(\tl(i)-\tk(i))})$ rows, and $M'_i$ has $m_i r_i$ rows, where
$r_i = 2^{r-i}$. Therefore, the number of rows of $M$ is
\[
\sum_{i=0}^r r_i m_i = \sum_{i=0}^r 2^{t+\log u' + r+1} m_i = 2^t d
\sum_{i=0}^r O_u(2^{u(\tl(i)-\tk(i))}).
\]

Let $S, Z \subseteq \zo^\tn$ respectively denote any choice of a
critical set and zero set of size at most $d$, where $|Z| \leq |S|$,
and choose an integer $i \geq 0$ so that $2^{i-1} u' \leq |S| \leq 2^i
u'$.  Arbitrarily grow the two sets $S$ and $Z$ to possibly larger,
and disjoint, sets $S' \supseteq S$ and $Z' \supseteq Z$ such that
$|S'| = |Z'| = 2^i u'$ (for simplicity we have assumed that $d \leq
n/2$).  Our goal is to show that there are ``many'' rows of the
matrix $\cM_i$ (in Construction~\ref{constr:reg}) that $u$-satisfy $S$ and
$Z$.

Let $\tk := \tk(i) = \log u' + i+1$, $\tl := \tl(i)$, and denote by
$G_1, G_2, G_3$ the bipartite graphs used by the instantiation of
Construction~\ref{constr:main} that outputs $\cM_i$. Thus we need to
show that ``many'' right vertices of $G_3$ 
are each connected to exactly $u$ of the vertices in $S$ and none of
those in $Z$.

Consider the uniform distribution $\cX$ on the set $S' \cup Z'$, which
has min-entropy $\log u' + i+1$.  By an averaging argument, since the
condenser $f_i$ is strong, for more than a $p$ fraction of the choices
of the seed $y \in \zo^\tee$ (call them \emph{good seeds}), the
distribution $\cZ_y := f_i(\cX, y)$ is $\eps/(1-p)$-close (in
particular, $(1/32)$-close) to a distribution with min-entropy $\log u'
+ i+1$.

Fix any good seed $y \in \zo^\tee$.  Let $G=(\zo^\tn, \zo^\tl, E)$
denote a bipartite graph representation of $f_i$, where each left
vertex $x \in \zo^\tn$ is connected to $f_i(x,y)$ on the right. Denote
by $\Gamma_y(S' \cup Z')$ the right vertices of $G$ corresponding to
the neighborhood of the set of left vertices picked by $S' \cup
Z'$. Note that $\Gamma_y(S' \cup Z') = \supp(\cZ_y)$.  Using
Proposition~\ref{prop:flatmap}, we see that since $\cZ_y$ is
$(1/32)$-close to having min-entropy $\log(|S' \cup Z'|)$, there are at
least $(7/8) |S' \cup Z'|$ vertices in $\Gamma(S' \cup Z')$ that are
each connected to exactly one left vertex in $S' \cup Z'$.  Since $|S|
\geq |S' \cup Z'| / 4$, this implies that at least $|S' \cup Z'|/8$
vertices in $\Gamma(S' \cup Z')$ (call them $\Gamma'_y$) are connected
to exactly one left vertex in $S$ and no other vertex in $S' \cup Z'$.
In particular we get that $|\Gamma'_y| \geq 2^{k-3}$.

Now, in $G_1$, let $T_y$ be the set of left vertices corresponding to
$\Gamma'_y$ (regarding the left vertices of $G_1$ in one-to-one
correspondence with the right vertices of $G$).  The number of edges
going out of $T_y$ in $G_1$ is $d_\ell |T_y| \geq u 2^k$. Therefore,
as the number of the right vertices of $G_1$ is $2^k$, there must be
at least one right vertex that is connected to at least $u$ vertices
in $T_y$.  Moreover, a counting argument shows that the number of
right vertices connected to $u$ or more vertices in $T_y$ is at
least $2^{k-\tl} 2^k/(10u)$. 

Observe that in construction of
$G_2$ from $G_1$, any right vertex of $G_1$ is replicated $\binom{d_r}{u}$
times, one for each $u$-subset of its neighbors. Therefore, for a right
vertex of $G_1$ that is connected to \emph{at least} $u$ left vertices in
$T_y$, one or more of its copies in $G_2$ must be connected to \emph{exactly}
$u$ vertex in $T_y$ (among the left vertices of $G_2$) and no other vertex (since the right degree of 
$G_2$ is equal to $u$).

Define $\gamma' := \max\{1, 2^{k-\tl} 2^k/(10u)\}$. From the
previous argument we know that,
looking at $T_y$ as a set of left vertices of $G_2$, there are at
least $\gamma'$ right vertices on the neighborhood of $T_y$ in $G_2$
that are connected to exactly $u$ of the vertices in $T_y$ and none of
the left vertices outside $T_y$. Letting $v_y$ be any such vertex,
this implies that the vertex $(y,v_y) \in V_3$ on the right part of
$G_3$ is connected to exactly $u$ of the vertices in $S$, and none of
the vertices in $Z$.  Since the argument holds for every good seed
$y$, the number of such vertices is at least the number of good seeds,
which is more than $p \gamma' 2^t$. Since the rows of the matrix $m_i$
are repeated $r_i = 2^{r-i}$ times in $M$, we conclude that $M$ has at
least $p \gamma' 2^{t+r-i} \geq p \gamma 2^t$ rows that $u$-satisfy
$S$ and $Z$, and the claim follows.


\end{Proof}

\subsection{Instantiations} \label{sec:instan}

We now instantiate the result obtained in Theorem~\ref{thm:regular} by various
choices of the family of lossless condensers. The crucial factors that influence the
number of measurements are the seed length and the output length of
the condenser.

Non-constructively, it can be shown that strong $(\tk, \eps)$ lossless
condensers with input length $\tn$, seed length $\tee = \log \tn +
\log(1/\eps) + O(1)$, and output length $\tl = \tk+\log(1/\eps)+ O(1)$
exist, and moreover, almost matching lower bounds are known
\cite{ref:CRVW02}.  In fact, the optimal parameters can be achieved by a
random function with overwhelming probability.  In this work, we 
consider two important explicit constructions of lossless condensers.
Namely, one based on ``zig-zag products'' due to Capalbo et al.~\cite{ref:CRVW02}
and another, coding theoretic, construction due to Guruswami et al.~\cite{ref:GUV07}.

\begin{thm} \cite{ref:CRVW02} \label{thm:CRVW} For every $\tk \leq \tn
  \in \N$, $\eps > 0$ there is an explicit lossless $(\tk, \eps)$
  condenser with seed length 
  $O(\log^3 (\tn/\eps))$ and output length
  $\tk+\log(1/\eps)+O(1)$. 
\end{thm}

\begin{thm} \cite{ref:GUV07} \label{thm:GUVcond} For all constants
  $\alpha \in (0,1)$ and every $\tk \leq \tn \in \N$, $\eps > 0$ there
  is an explicit strong lossless $(\tk, \eps)$ condenser with seed
  length $\tee=(1+1/\alpha) \log (\tn \tk/\eps) + O(1)$ and output
  length $\tl=\tee+(1+\alpha)\tk$. 
\end{thm}
As a result, we use Theorem~\ref{thm:regular} with the above condensers to obtain the following.

\begin{thm} \label{thm:instan} Let $u > 0$ be fixed, and $p \in [0,1)$
  be a real parameter. Then for integer parameters $d, n \in \N$ where
  $u \leq d \leq n$,

  \begin{enumerate}
  \item Using an optimal lossless condenser in
    Construction~\ref{constr:reg} results in an $m_1 \times n$ matrix
    $\cM_1$ that is $(d,e_1; u)$-regular, where $m_1 = O(d (\log n)
    (\log d) / (1-p)^{u+1})$ and $e_1 = \Omega(p d \log n)$,
 
  \item Using the lossless condenser of Theorem~\ref{thm:CRVW} in
    Construction~\ref{constr:reg} results in an $m_2 \times n$ matrix
    $\cM_2$ that is $(d,e_2; u)$-regular, where $m_2 = O(T_2 d (\log
    d) / (1-p)^u)$ for some $T_2 = \exp(O(\log^3((\log n)/(1-p)))) =
    \qpoly(\log n)$, and $e_2 = \Omega(pd T_2 (1-p))$.
 
  \item Let $\beta > 0$ be any fixed constant. Then
    Construction~\ref{constr:reg} can be instantiated using the
    lossless condenser of Theorem~\ref{thm:GUVcond} so that we obtain
    an $m_3 \times n$ matrix $\cM_3$ that is $(d, e_3; u)$-regular,
    where $m_3 = O(T_3^{1+u} d^{1+\beta} (\log d))$ for $T_3 := ((\log
    n) (\log d)/(1-p))^{1+u/\beta} = \poly(\log n, \log d)$, and $e_3 =
    \Omega(p \max\{ T_3, d^{1-\beta/u}\})$.
  \end{enumerate}
\end{thm}

\begin{Proof}
First we show the claim for $M_1$. In this case, we take each $f_i$ in
Construction~\ref{constr:reg} to be an optimal lossless condenser
satisfying the (non-constructive) bounds obtained in\footnote{This result is similar
  in spirit to the probabilistic argument used in
  \cite{ref:lowerbounds} for showing the existence of good
  extractors.} \cite{ref:CRVW02}.  Thus we have that $2^\tee =
O(\tn/\eps)=O(\log n/\eps)$, and for every $i = 0, \ldots, r$, we have
$2^{\tl(i)-\tk(i)} = O(1/\eps)$, where $\eps = O(1-p)$. Now we 
apply Theorem~\ref{thm:regular} to obtain the desired bounds (and in
particular, $\gamma = \Omega(\eps d)$).
 
Similarly, for the construction of $\cM_2$ we set up each $f_i$ with
the explicit construction of condensers in 
Theorem~\ref{thm:CRVW} for min-entropy $\tk(i)$. In this case, the
maximum required seed length is $t = O(\log^3(\tn/\eps))$, and we let
$T_2 := 2^t = \exp(O(\log^3((\log n)/(1-p))))$. Moreover, for every $i
= 0, \ldots, r$, we have $2^{\tl(i)-\tk(i)} = O(1/\eps)$. Plugging
these parameters in Theorem~\ref{thm:regular} gives $\gamma =
\Omega(\eps d)$ and the bounds on $m_2$ and $e_2$ follow.
 
Finally, for $\cM_3$ we use Theorem~\ref{thm:GUVcond} with $\alpha :=
\beta/u$.  Thus the maximum seed length becomes $t = (1+u/\beta)
\log(\tn (\log d)/(1-p)) + O(1)$, and for every $i = 0, \ldots, r$, we have
$\tl(i)-\tk(i) = O(t+\beta (\log d)/u)$.  Clearly, $T_3 = \Theta(2^t)$,
and thus (using Theorem~\ref{thm:regular}) the number of measurements
becomes $m_3 = T^{1+u} d^{1+\beta} (\log d)$. Moreover, we get $\gamma
= \max\{1, \Omega(d^{1-\beta/u} /T )\}$, which gives $e_3 =
\Omega(pT\gamma) = p \max\{T, d^{1-\beta/u}\}$, as claimed.
\end{Proof}

By combining this result with
Lemma~\ref{lem:repdisj} using any explicit construction of classical disjunct
matrices, we obtain threshold $(d,e;u)$-disjunct matrices that can be
used in the threshold model with any fixed threshold, sparsity $d$,
and error tolerance $\lfloor e/2 \rfloor$. In particular, using the coding-theoretic explicit
construction of nearly optimal classical disjunct matrices from codes
on the Gilbert-Varshamov bound \cite{ref:PR08}
(Theorem~\ref{thm:GVdisjunct} in the appendix), we  
obtain threshold $(d,e;u)$-disjunct matrices with $m=O(m' d^2 (\log n)/(1-p)^2)$ rows
and error tolerance $e = \Omega(e' p d (\log n)/(1-p))$, where $m'$ and $e'$ are respectively
the number of rows and error tolerance of any of the regular matrices
obtained in Theorem~\ref{thm:instan}. 
We note that in all cases, the final dependence on the sparsity parameter $d$ is, roughly,
$O(d^3)$ which has an exponent independent of the threshold $u$.
Rows M7--M9 of Table \ref{tab:params}
summarize the obtained parameters for the general case (with arbitrary gaps).
We see that, when $d$ is not negligibly small (e.g., $d=n^{1/10}$), the bounds
obtained by our explicit constructions are significantly better than those 
offered by strongly disjunct matrices.

%
%

\section{The case with positive gaps} \label{sec:gap}


In preceding sections we have focused on the case where $g=0$. 
However, in this section we observe that all the techniques that we have developed 
in this work can be extended
to the positive-gap case in a straightforward way. The main
observations are listed below.
Recall from \cite{ref:thresh1} that in the positive-gap case, we can only hope
to distinguish between distinct $d$-sparse vectors $x$ and $x'$ where at least
one has support size $u$ or more and either $|\supp(x) \setminus \supp(x')| > g$ or
$|\supp(x') \setminus \supp(x)| > g$. We will call any pair of such vectors
\emph{distinguishable}. Moreover, we naturally extend the Definition~\ref{def:thresholdDesign}
of threshold designs to the positive-gap case as follows.

\begin{defn}[Definition~\ref{def:thresholdDesign}, generalized] \label{def:thresholdDesignGeneral}
Let $n \geq d \geq u > 0$ and $g \in [0,u)$, and $e \geq 0$ be integer parameters, and define $\ell := u-g-1$.
A Boolean matrix $M$ with $n$ columns is said to be a \emph{$(d, e; u, g)$-threshold design}
if for every $d$-sparse $x, x' \in \zo^n$ of Hamming weight $u$ or more 
such that $|\supp(x) \setminus \supp(x')| > g$,
every $y \in M[x]_{\ell,u}$ and every $y' \in M[x']_{\ell,u}$, we have
$\dist(y, y') > e$. 
\end{defn}


\subsection{Generalized threshold disjunct matrices}
For the positive-gap case, 
Definition~\ref{def:regularMatrix} of threshold disjunct matrices can be adapted to allow more
  than one distinguished column in disjunct matrices. In particular,
  in general we may require the matrix $\cM$ to have more than $e$ rows that
  $u$-satisfy every choice of a critical set $S$, a zero set $Z$, and
  any set of $g+1$ designated columns $I \subseteq S$ (at which all entries
  of the corresponding rows must be $1$).  Denote this generalized
  notion by threshold $(d,e;u,g)$-disjunct matrices.  It is straightforward to
  extend the arguments of Lemma~\ref{lem:disjunct} to show that the
  generalized notion of threshold $(d,e;u,g)$-disjunct matrices suffices to
  capture non-adaptive threshold group testing with
  upper threshold $u$ and gap $g$. More precisely, the generalized definitions
  of threshold disjunct and regular matrices are as follows.

\begin{defn}[Definition~\ref{def:regularMatrix}, generalized] \label{def:regularMatrixGeneral} 
Let $n, d, e, u, g$ be non-negative integers where
$g < u \leq d \leq n$.
A Boolean matrix $\cM$ with $n$
  columns is called threshold $(d,e;u,g)$-disjunct if for every subset of columns
  $S \subseteq [n]$ (called the \emph{critical set}), every $Z
  \subseteq [n]$ (called the \emph{zero set}) such that $u \leq |S|
  \leq d$, $|Z| \leq |S|$, $S \cap Z = \emptyset$, and every set
  $I \subseteq S$ of $g+1$ distinguished columns ($|I|=g+1$), there are more than
  $e$ rows of $\cM$ that $u$-satisfy $S$ and $Z$ and moreover, $M|_I$ has
  all ones at those columns. 
  Moreover, $\cM$ is called $(d,e;u,g)$-regular if for every choice of the
  critical and zero sets $S,Z \subseteq [n]$ with $|Z| \leq |S| + g$,
  there is a set of more than $e$ rows of $\cM$ that $(u-g)$-satisfy
  $S$ and $Z$.
\end{defn}

Note the slight difference between the notion of regular matrices above compared
to Definition~\ref{def:regularMatrix}, namely, that the zero set $Z$ can now be slightly
larger than the critical set $S$ (by at most $u$), and that the matrix is now required 
to $(u-g)$-satisfy (as opposed to $u$-satisfy) every choice of $S$ and $Z$.
The two notions coincide for $g=0$. In general, the difference between the two notions
of regular matrices is negligible as long as the parameter $g$ remains small.
In particular, it is straightforward to verify that all our results about the
construction of regular matrices in the gap-free case (Constructions \ref{constr:prob}~and~\ref{constr:reg})
as well as the obtained bounds (Lemma~\ref{lem:probDisjunct}, Theorem~\ref{thm:regular}
and Theorem~\ref{thm:instan}) hold for the generalized notion of regular matrices with
only a slight effect on the hidden terms that only depend on the threshold parameter $u$.
We will see, however, that the generalized notion of threshold disjunct matrices is
stronger than Definition~\ref{def:regularMatrix} and the extra requirements may substantially affect the bounds
(but not the construction techniques).

Below we show that the generalized notion of threshold disjunct matrices 
suffices for construction of threshold designs for the positive-gap case.

\begin{lem}[Lemma~\ref{lem:disjunct}, generalized] \label{lem:disjunctGeneral} Let $\cM$ be an $m \times n$ Boolean
  matrix that is threshold $(d,e;u,g)$-disjunct, and define $\ell := u-g-1$. Then for every distinguishable
  $d$-sparse vectors $x, x' \in \zo^n$, each having support size $u$ or more and 
  such that
    $|\supp(x) \setminus \supp(x')| > g$ and
  $\wgt(x) \geq |\supp(x') \setminus \supp(x)|$, the following holds.
  Let $y \in \cM[x]_{\ell,u}$ and $y' \in \cM[x']_{\ell,u}$. Then,
  \begin{equation} \label{eqn:distinGeneralized} |\supp(y) \setminus
    \supp(y')| > e.
  \end{equation}
  Moreover, $M$ is a $(d,e;u,g)$-threshold design.
  Conversely, if $\cM$ satisfies \eqref{eqn:distinGeneralized} for every choice
  of $x$, $x'$, $y$, $y'$ as above, it must be threshold $(\lfloor d/2
  \rfloor,e;u,g)$-disjunct (assuming $n > d+g$).
\end{lem}

\begin{Proof}
  First, suppose that $\cM$ is threshold $(d,e;u,g)$-disjunct, and let $y \in
  \cM[x]_{\ell,u}$ and $y' \in \cM[x']_{\ell,u}$ be arbitrarily chosen.
  Take any $I \subseteq \supp(x) \setminus
  \supp(x')$ of size $g+1$, and let $S := \supp(x)$ and $Z := \supp(x') \setminus
  \supp(x)$.  Note that $|S| \leq d$ and by assumption, we have $|Z|
  \leq |S|$.
  Now, Definition~\ref{def:regularMatrixGeneral} implies that there is a set
  $E$ of more than $e$ rows of $M$ that $u$-satisfy $I$ as the set of
  distinguished columns, $S$ as the critical set and $Z$ as the zero
  set. Thus for every $j \in E$, the $j$th row of $\cM$ restricted to
  the columns chosen by $\supp(x)$ must have weight exactly $u$, while
  its weight on $\supp(x')$ is at most $u-g-1 = \ell$. Therefore, $y(j) = 1$
  and $y'(j) = 0$ for more than $e$ choices of $j$.

  The claim that $\cM$ is a $(d,e;u,g)$-threshold design follows from
  the above argument combined with the observation that, given
  any two $d$-sparse distinguishable vectors, having
  Hamming weight $u$ or more, at least 
  one of their two possible orderings satisfies the conditions required by the lemma.

  For the converse, consider any choice of a set of distinguished columns $I
  \subseteq [n]$ with $|I| = g+1$, a critical set $S \subseteq [n]$ containing $I$ (such that
  $|S| \geq u$), and a zero set $Z \subseteq [n]$ where $|Z| \leq
  |S|$.  Define $d$-sparse Boolean vectors $x, x' \in \zo^n$ so that
  $\supp(x) := S$ and $\supp(x') := Z \cup (S \setminus I)$.  
  We note that $\wgt(x) = |\supp(x)| \geq u$ and also, without loss of generality, $\wgt(x') \geq u$ 
  (if the latter is not the case, we can simply enlarge $Z$ by arbitrarily adding up to $g+1$
  elements outside the support of $S$ to it and observe that is suffices to show the
  claim for the larger $Z$).
  
  Let $y := M[x]_{\ell+1}$ and $y' := M[x']_u$, and observe that
$y, y' \in M[x]_{\ell,u}$. 
Moreover, let $E := \supp(y) \setminus
  \supp(y')$. By assumption we know that $|E| > e$. Take any $j \in
  E$. Since $y(j) = 1$ and $y'(j) = 0$, we get that the $j$th row of
  $\cM$ restricted to the columns picked by $Z \cup (S \setminus I)$
  must have weight at most $\ell=u-(g+1)$, whereas it must have weight at least
  $u$ when restricted to $S$.  As the sets $I, S \setminus I$,
  and $Z$ are disjoint and $|I| = g+1$, this can hold only if 
  the $j$th row of $\cM$ restricted to the columns picked by
  $S$, $Z$, and $I$ has weights exactly $u$, $0$, and $g+1$, respectively.
  Hence, this
  row (as well as all the rows of $\cM$ picked by $E$) must $u$-satisfy
  $I, S$, and $Z$, confirming that $\cM$ is threshold $(\lfloor d/2
  \rfloor,e;u,g)$-disjunct.
\end{Proof}


\subsection{Strongly disjunct versus threshold disjunct matrices}


The following proposition directly follows from the definitions, and relates strongly disjunct
matrices to generalized threshold disjunct matrices.

\begin{prop} \label{prop:strongVSthreshold}
Let $n, d, e, u, g$ be non-negative integers where
$g < u \leq d \leq n-(d+g+1)$.
Suppose that $\cM$ and $\cM'$ are binary $m \times n$ matrices, where
$\cM$ is threshold $(d,e;u,g)$-disjunct and $\cM'$ is strongly $(2d,e;u)$-disjunct.
Then, $\cM$ is strongly $(d,e;g+1)$-disjunct and $\cM'$ is threshold $(d,e;u,g)$-disjunct.
\end{prop}

\begin{proof}
First, we verify the conditions of Definition~\ref{def:strongDisjunct} for $\cM$.
Consider any pair of disjoint sets $I,Z \subseteq [n]$ where $|I|=g+1$ and $|Z| \leq d$.
Let $S \subseteq [n]$ be any set of size $d$ containing $I$ and disjoint from $Z$. Note
that $|Z| \leq |S|$.
From Definition~\ref{def:regularMatrixGeneral} (with the
critical set $S$, zero set $Z$, and distinguished set $I$), there is a set of more than $e$
rows of $\cM$ at which $\cM|_Z$ is all zeros and $\cM|_I$ is all ones. In
other words, denoting the $i$th column of $\cM$ by $C_i$, we have that
  \[
  |\cap_{i \in I} \supp(C_i) \setminus \cup_{i \in Z} \supp(C_i) | > e,
  \]
as required by Definition~\ref{def:strongDisjunct}.

Now consider the matrix $M'$ and any choice of a $I,S,Z$ as in Definition~\ref{def:regularMatrixGeneral}.
Let $J \subseteq S$ be any subset of $S$ of size $u$ that contains $I$,
and $S' := Z \cup (S \setminus J)$. Note that $|S'| \leq |S| + |Z| \leq 2d$.
Now from Definition~\ref{def:strongDisjunct} of strongly disjunct matrices, 
we know that
  \[
  |\cap_{i \in J} \supp(C_i) \setminus \cup_{i \in S'} \supp(C_i) | > e.
  \]
In other words, there is a set of more than $e$ rows of $\cM'$ at which
$\cM'|_I$ is all ones, $\cM'|_S$ has weight exactly $u$, and $\cM'|_Z$ is
all zeros, as required by Definition~\ref{def:regularMatrixGeneral}.
\end{proof}

The special case $u=g+1$ in the above proposition is particularly interesting. 
A chain of reductions between strongly disjunct and threshold disjunct matrices
in this case implied by the above result is schematically shown below.

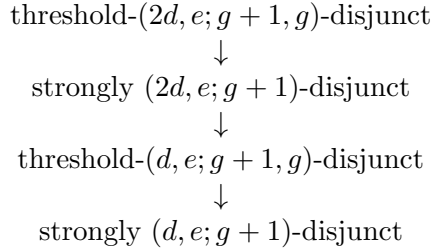
\[
\begin{array}{c}
\text{threshold-$(2d,e;g+1,g)$-disjunct} \\
\downarrow \\
\text{strongly $(2d,e;g+1)$-disjunct} \\
\downarrow \\
\text{threshold-$(d,e;g+1,g)$-disjunct} \\
\downarrow \\
\text{strongly $(d,e;g+1)$-disjunct} 
\end{array}
\]

Therefore, when the upper threshold $u$ is more than the gap parameter $g$ by one (equivalently,
when the lower threshold $\ell$ is zero), the two notions of threshold disjunct matrices
and strongly disjunct matrices become equivalent 
up to a multiplicative factor in the sparsity parameter $d$.
As discussed in Section~\ref{sec:disjunct}, almost matching lower bounds
and upper bounds on the number of rows $m$ achievable by a strongly $(d,e;g+1)$-disjunct matrix are known.
Asymptotically, the number of rows must always satisfy $m = \Omega(d^{g+2} \log_d n + e d^{g+1})$ and 
moreover, a probabilistic construction achieves $m = O(d^{g+2} \log(n/d)$ and $e = \Omega(d \log(n/d))$
with probability $1-o(1)$ (see Table~\ref{tab:params}). As a result, the upper and lower bounds
on the number of rows of strongly disjunct and threshold disjunct matrices become equivalent up 
multiplicative constants when the lower threshold $\ell$ is zero.

Proposition~\ref{prop:strongVSthreshold} asserts that the notion of strongly disjunct matrices
is in general stronger than threshold disjunct matrices. As we will see below,
the former becomes strictly stronger when $\ell > 0$. As the lower threshold $\ell$ becomes
larger, the discrepancy between the number of rows achievable by threshold disjunct matrices 
and strongly disjunct matrices becomes more significant (see Table~\ref{tab:params}).
  
\subsection{Probabilistic upper bounds}

As pointed out after Definition~\ref{def:regularMatrixGeneral}, the generalized
definition of regular matrices may affect the bounds obtained by our
probabilistic and explicit constructions (Constructions \ref{constr:prob} and \ref{constr:reg})
only by hidden factors depending on $u$ (essentially without any change in the proofs).
For the case of generalized disjunct matrices, however, the bounds may
substantially change depending on the gap parameter $g$.

Below we generalize Lemma~\ref{lem:probDisjunct} for the case of threshold disjunct
matrices and show that
  Construction~\ref{constr:prob} results
  in a threshold $(d, \Omega_u(pd \log(n/d)/(1-p)^2); u,g)$-disjunct matrix (with probability $1-o(1)$) if the number of
  measurements is increased by a factor $O(d^g)$. More precisely, we can now
  show the following lemma.

\begin{lem}[Lemma~\ref{lem:probDisjunct}, generalized] \label{lem:probDisjunctGeneral} For every $p \in [0,1)$ and
  integer parameters $u > g \geq 0$, Construction~\ref{constr:prob} with
  $m' = O_{u}(d^{g+2} \log (n/d)/(1-p)^2)$ 
  outputs a threshold $(d, \Omega_{u}(pm'/d^{g+1});u,g)$-disjunct matrix with probability $1-o(1)$.
\end{lem}

 \begin{Proof}
The proof essentially follows along the same lines as the proof of 
Lemma~\ref{lem:probDisjunct}. The difference, compared to the case
$g=0$ covered by Lemma~\ref{lem:probDisjunct}, is that we have a set $I$ of
distinguished columns $I \subseteq [n]$ in Definition~\ref{def:regularMatrixGeneral}
where $|I| = g+1$ and the random vector $\rv{w}$ in the proof of 
 Lemma~\ref{lem:probDisjunct} must have ones at all positions picked by $I$.
With this requirement, the lower bound on the success probability $q$
in \eqref{eqn:succesProb} becomes $c = \Omega_u(1/d^{g+1})$.
The rest of the proof remains unchanged except for the new lower bound on $c$,
which makes the error tolerance parameter $e$ in the proof 
lower bounded by $\Omega(pm'/d^{g+1})$, while increasing the parameter $m'$ to 
a quantity upper bounded by $O_{u}(d^{g+2} \log (n/d)/(1-p)^2)$.
\end{Proof}

\subsection{The direct product}
Lemma \ref{lem:repdisj} can be extended to positive gaps as follows.

  
\begin{lem}[Lemma~\ref{lem:repdisj}, generalized] \label{lem:repdisjGeneralized} 
Suppose that $\cM_1$ is a
  $(d,e_1;u,g+1)$-regular and $\cM_2$ is a strongly $(2d,e_2;g+1)$-disjunct matrix.
  Then $\cM_1 \rep \cM_2$ is a threshold $(d, (e_1+1) (e_2+1) -1; u,g)$-disjunct
  matrix. 
\end{lem}

\begin{Proof}
Let $\ell := u-g-1$ and $\cM := \cM_1 \rep \cM_2$.
Towards verifying that $\cM$ satisfies the requirements of Definition~\ref{def:regularMatrixGeneral},
consider a set $I \subseteq [n]$ of distinguished columns of $\cM$, where $n$ is the
number of columns of the matrices and $|I|=g+1$, in addition to critical and zero sets $S, Z \subseteq [n]$
as in Definition~\ref{def:regularMatrixGeneral} satisfying $|Z| \leq |S|$. Index the rows of $\cM$ naturally by
the elements of $[m_1] \times [m_2]$, where $m_1$ and $m_2$ are the number of rows of $\cM_1$ and
$\cM_2$, respectively, and the $(i,j)$th row of $\cM$ is the bitwise disjunction of the
$i$th row of $\cM_1$ and the $j$th row of $\cM_2$.

Let $S' := S \setminus I$ and $Z' := Z \cup (S \setminus I)$.
Observe that $|Z| \leq |S| \leq |S'|+g+1 = |S|+|I|$ and $|Z'| \leq 2d$. 
From Definition~\ref{def:regularMatrixGeneral}, there is a set
$E_1 \subseteq [m_1]$ of size more than $e_1$ such that
$\cM_1|_{S'}$ has weight exactly $\ell = u-|I|$, and $\cM_2|_{Z}$ is all zeros. 
Moreover, there is a set $E_2 \subseteq [m_2]$ of size more
than $e_2$ at which $\cM_2|_I$ has all ones and $\cM_2|_{Z'}$ has all zeros. 
This means that, at all rows corresponding to $E_1 \times E_2$,
the product matrix $\cM$ has weight exactly $\ell+|I| = u$ at
positions corresponding to $S$ and all zeros at positions corresponding to $Z$.
Therefore, $\cM$ indeed $u$-satisfies any choice of the sets $I$, $S$, $Z$
at more than $(e_1+1)(e_2+1)-1$ rows.
\end{Proof}  
  
  Consequently, using the coding-theoretic construction of
  strongly disjunct matrices described in Section~\ref{app:strongly},
  our explicit constructions of threshold $(d,e;u)$-disjunct matrices obtained 
  in Section~\ref{sec:constr} can be
  extended to the positive gap model at the cost of a factor $O(d^g)$ increase
  in the number of measurements. The results from combining the above
  lemma with various constructions of regular and strongly disjunct matrices
  are summarized in Table~\ref{tab:params}.  


\subsection{Lower bounds}

We now extend the lower bounds proved in Section~\ref{sec:lower}
to the positive-gap case, and show that the optimal exponent of
$d$ in the number of measurements is $g+1$. 

The lower bound on the number of rows of threshold disjunct matrices
is an immediate consequence of Proposition~\ref{prop:strongVSthreshold}.

\begin{thm} \label{thm:lowerBoundThresholdDisjunctGeneral}
For every integer $d > 0$ there is an $n_0 > 0$ such that the following holds.
For any $n \geq n_0$, let $M$ be an $m \times n$ threshold $(d,e;u,g)$-disjunct
matrix. Then, for some absolute constant $c > 0$,
\[
m \geq 0.7 c \frac{\binom{g+d+1}{g+1} (g+d+1)}{\log \binom{g+d+1}{g+1}} \log n + 
0.5 c \binom{g+d+1}{g+1} e 
= \Omega\big( (d/g)^{g+2} \log_d n + 
(d/g)^{g+1} e \big).
\]
\end{thm}

\begin{proof}
Immediate from combining Proposition~\ref{prop:strongVSthreshold}
and Theorem~2.19 of \cite{ref:SW04} that proves a lower bound on the
number of rows of strongly disjunct matrices. The asymptotic
simplification is straightforward.
\end{proof}

In order to lower bound the number of measurements in a threshold design, we first
generalize Lemma~\ref{lem:necessity} as follows.

\begin{lem} \label{lem:necessityGeneral}
Let $M$ be a $(d+g,e;u,g-1)$-threshold design, and $\ell := u-(g-1)-1 = u-g$. be the
lower threshold 
Then $M$ satisfies the following property:

``For every $S \subseteq [n]$ such that $|S| = d$ and every $T \subseteq [n] \setminus S$
such that $|T| = g$,
there are more than $e$ rows of $M$ at which $M|_T$ consists of all ones and
moreover in those rows, $M|_S$ has weight exactly $\ell$.''
\end{lem}

\begin{proof}
%
Let $D := d+g$ be the sparsity parameter in the threshold model that
$M$ is designed for.
In order to verify the claimed property for a given choice of $S$ and $T$, 
consider the $D$-sparse vectors $x, x' \in \zo^n$ such that
$\supp(x) = S \cup T$ and $\supp(x') = S$.
Let $y := M[x]_{\ell+1}$ and $y' := M[x']_u$, and observe that
$y, y' \in M[x]_{\ell,u}$. Also, since $x'$ is point-wise less than
or equal to $x$, or in symbols $x' \preceq x$, by monotonicity it also
follows that $y' \preceq y$.

Thus, we know from the assumption that there are more than $e$ positions at which $y$ and
$y'$ are different. Let $j$ be any such position. In particular, 
we know that $y(j) = 1$ and $y'(j) = 0$. 
Therefore, by Definition~\ref{def:thresholdDesignGeneral} and
the way that the threshold model is defined, the submatrix
$M|_{\supp(x')}$ must have weight at most $\ell$ at the $j$th row and
$M|_{\supp(x)}$ must have weight at least $u$ at the $j$th row.
Since the support of $x$ is chosen to be larger than the support of $x'$
at exactly $g$ positions, and $g = u-\ell$, 
the only possibility is to have
$M|_{\supp(x)}$ (that is, $M|_{S \cup T}$) with weight \emph{exactly} $u$ at the $j$th row and
$M|_{\supp(x)\setminus \supp(x')}$ (that is, $M|_{T}$) having all ones at the $j$th row.
In turn, this implies that $M|_{\supp(x')}$ (that is, $M|_{S}$) must have
weight exactly $\ell$ at that row.

This concludes proof, since the argument holds for every possible choice of the
distinguishing entry $j$.
\end{proof}

The following theorem is the analogous version of Theorem~\ref{thm:lowerboundGapfree}
for the positive-gap case.

\begin{thm} \label{thm:lowerboundGeneral}
For every integer $d > 0$ there is an $n_0 > 0$ such that the following holds.
For any $n \geq n_0$, let $M$ be an $m \times n$ Boolean matrix that satisfies the
property quoted in Lemma~\ref{lem:necessityGeneral}. Then,
\[
m = \Omega\Big( \frac{m'}{\log m'} + 
\big(\frac{d}{\ell+1}\big)^2 \log_d n + \frac{de}{(\ell+1)^2}\Big), \text{ where } 
m' := \frac{(d/\ell g)^{g+1} \log_d n + (d/\ell g)^g e}{(g+\ell)\log n}.
\]
In particular, when $\ell, g$ are absolute constants, we have
\[
m = \Omega_{\ell,g}\Big( \frac{1}{\log d+\log(e+2)} \cdot \big( \frac{d^{g+1}}{\log d} + \frac{d^g e}{\log n} \big)  + d^2 \log_d n + de \Big).
\]
\end{thm}

\begin{proof}
First, observe that the property quoted in Lemma~\ref{lem:necessityGeneral}
is stronger than the property quote in Lemma~\ref{lem:necessity}. Thus,
the lower bound of Theorem~\ref{thm:lowerboundGapfree} holds; namely, we have
\[
m = \Omega\Big( \big(\frac{d}{\ell+1}\big)^2 \log_d n + \frac{de}{(\ell+1)^2}\Big).
\]
Therefore, it suffices to show that $m = \Omega( m'/\log m')$.
Given the matrix $M$, we will use a ``random resampling method'' to
create a strongly disjunct matrix out of $M$, and will then use the
known lower bounds related to strongly disjunct matrices.

Given a vector $v \in \zo^n$, a \emph{resampling} of $v$ is a
random vector $v' \in \zo^n$ defined in the following way:
For ever $i \in [n]$, if $v(i) = 0$, then we set $v'(i) = 0$.
Otherwise, we independently set $v'(i) = 1$ with probability
$1/\ell$ and zero with the remaining probability.

Let $t > 0$ be an integer parameter to be determined later. 
Let $M_1, \ldots M_t$ be random $m \times n$
Boolean matrices, such that each $M_i$ is obtained from $M$
by independently resampling each row of the matrix.
Also, define the $mt \times n$ Boolean matrix $M'$ by 
stacking $M_1, \ldots, M_t$ on top of one another. We will
argue that, if $t$ is chosen sufficiently large, there is
a nonzero probability that $M'$ becomes a strongly disjunct
matrix. Thus, there is a fixing of the resampling randomness
that indeed makes $M'$ strongly disjunct, which then allows us to
obtain the desired lower bound.

Consider any triple $(j,T,W)$ where $j \in [m]$, $T, W \subseteq [n]$
such that $|T| = g$, $|W| = \ell$, $T \cap W = \emptyset$, and moreover,
the $j$th row of $M$ has all-ones at the columns picked by $T$ and $W$.
We say that the triple \emph{survives} in $M_i$ when on the
$j$th row of $M_i$, the columns picked by $T$ all contain ones
and those picked by $W$ all contain zeros.
The probability of this happening is exactly 
\[
p = (1 - 1/\ell)^\ell \cdot 1/\ell^g \geq c/\ell^g,
\]
for some absolute constant $c > 0$.
The probability that the triple does not survive in any
of $M_1, \ldots, M_t$ is
\[
(1-p)^t \leq (1-c/\ell^g)^t \leq C^{t/\ell^g},
\]
for some absolute constant $C \in (0,1)$. Combined with a union
bound on all choices of $(j,T,W)$, we deduce that the probability
that some triple does not survive in any of
of $M_1, \ldots, M_t$ is at most
\[
m n^{g+\ell} C^{t/\ell^g} = 2^{\log m + (g+\ell) \log n - (t/\ell^g) \log(1/C)},
\]
which is strictly less than $1$ for some large enough choice
of $t$, namely, for $t \geq t_0$ such that
\[
t_0 = O(\ell^g((g+\ell) \log n + \log m)).
\]
Now, pick $t := t_0$ and fix the resampling randomness so that
all triples $(j,T,W)$ survive. We claim that the matrix $M'$ 
is strongly $(d, e; g)$-disjunct. 

In order to verify the disjunctness property, consider any
choice of sets $S, T \subseteq [n]$ such that $|S| = d$ and
$|T| = g$. Let $J$ be the set of rows of $M$ where 
$M|_T$ has all-ones and $M|_S$ has Hamming weight $\ell$.
By the property quoted in Lemma~\ref{lem:necessityGeneral},
we know that $|J| > e$.

For any $j \in J$, we know that the $j$th row of $M|_S$ is supported
on some set $W \subseteq [n]$ of size $\ell$. We know,
on the other hand, that the triplet $(j,T,W)$ survives in some
$M_i$. Clearly, by the way we defined the survival property, 
this implies that $j$th row of $M_i$ (and thus, the corresponding
row in $M'$) contains all-ones at
columns picked by $T$ and all-zeros at columns picked by $S$.
Since this argument holds for any choice of $S$, $T$, and $j$, we conclude
that $M'$ is strongly $(d, e; g)$-disjunct.

The number of rows of $M'$ is $mt$. We can now apply
the known lower bounds on the number of rows of strongly disjunct
matrices in order to lower bound the number of rows of $M'$.
In particular, Theorem~2.19 of \cite{ref:SW04} implies that, 
for some absolute constant $c' > 0$, and whenever $n$ is sufficiently large
for the given parameter $d$, 
\begin{eqnarray*}
mt &\geq& 0.7 c' \frac{\binom{g+d}{g} (g+d)}{\log \binom{g+d}{g}} \log n + 
0.5 c' \binom{g+d}{g} e \\
&\geq& 0.7 c' \frac{(g+d)^{g+1} \log n}{\log(g+d) g^{g+1}} \log n + 
0.5 c' \frac{(g+d)^g e}{g^g}  \\
&=& \Omega\big( (d/g)^{g+1} \log_d n + 
(d/g)^g e \big).
\end{eqnarray*}
Now we substitute the chosen value of $t$ in the above bound to obtain
\begin{eqnarray*}
m &=& \Omega\Big( \frac{(d/\ell g)^{g+1} \log_d n + (d/\ell g)^g e}{(g+\ell)\log n + \log m}  \Big).
\end{eqnarray*}
Now if $n$ is sufficiently large for the given $d$, we can ensure that the
conditions of Proposition~\ref{prop:technical} in the appendix are satisfied,
and the above bound implies that
\[
m = \Omega( m'/\log m' ), \text{ where } 
m' := \frac{(d/\ell g)^{g+1} \log_d n + (d/\ell g)^g e}{(g+\ell)\log n},
\]
as claimed. The simplification when $\ell$ and $g$ are absolute constants is straightforward.
\end{proof}

Theorem~\ref{thm:lowerboundGeneral} combined with Lemma~\ref{lem:necessityGeneral}
implies the desired lower bound on the number of measurements of a threshold design.
The following corollary summarizes the simplified bounds for the 
case $e=0$.

\begin{coro} \label{coro:lowerboundGeneralSimplified}
For every integer $d > 0$ there is an $n_0 > 0$ such that the following holds.
For any $n \geq n_0$, let $M$ be an $m \times n$ Boolean matrix that is a
\emph{$(d, 0; u, g)$-threshold design}, for constants $u > g \geq 0$.
Then,
\[
m = \Omega_u \Big( \frac{d^{g+2}}{\log^2 d} + \frac{d^2 \log n}{\log d} \Big).
\] \qed
\end{coro}

\section{Strongly disjunct matrices from codes} \label{app:strongly}

A well known coding-theoretic construction of combinatorial designs,
and classical disjunct matrices is due to Kautz and
Singleton~\cite{ref:KS64}, which was further refined in several
subsequent works (such as \cite{ref:DMR00,ref:DMR00b}).

In this section we describe a construction of strongly disjunct
matrices (as in Definition~\ref{def:strongDisjunct}) which is a
straightforward extension of the original construction of Kautz and
Singleton.
Construction~\ref{constr:strong} explains the idea, which is analyzed
in Lemma~\ref{lem:ks} below.
In this section we use standard tools from the theory of error-correcting codes.
The interested reader is referred the standard texts in coding theory 
(e.g., the books by
MacWilliams and Sloane \cite{ref:MS}, van~Lint \cite{ref:vanl}, and
Roth \cite{ref:Roth}) for background.


\addtocounter{footnote}{1}
\footnotetext[\value{footnote}]{We use the standard coding-theoretic notation of 
$(\tn, \tk, \td)_q$ code for a $q$-ary code of length
  $\tn$, size $q^k$, and minimum distance at least $\td$.}

\begin{constr} 
\caption{Extension of Kautz-Singleton's method \cite{ref:KS64}.}

  \begin{itemize}
  \item {\it Given: } An $(\tn, \tk, \td)_q$ error-correcting code$^{\arabic{footnote}}$ 
  $\C \subseteq [q]^{\tn}$, and integer parameter $u > 0$.

  \item {\it Output: } An $m \times n$ Boolean matrix $\cM$, where $n
    = q^\tk$, and $m = \tn q^u$.

  \item {\it Construction: } First, consider the mapping
    $\varphi\colon [q] \to \zo^{q^u}$ from $q$-ary symbols to column
    vectors of length $q^u$ defined as follows.  Index the coordinates
    of the output vector by the $u$-tuples from the set $[q]^u$. Then
    $\varphi(x)$ has a $1$ at position $(a_1, \ldots, a_u)$ if and only if there
    is an $i \in [u]$ such that $a_i = x$. Arrange all codewords of
    $\C$ as columns of an $\tn \times q^\tk$ matrix $\cM'$ with
    entries from $[q]$. Then replace each entry $x$ of $\cM'$ with
    $\varphi(x)$ to obtain the output $m \times n$ matrix $\cM$.
  \end{itemize}
  \label{constr:strong}
\end{constr}

\begin{lem} \label{lem:ks} Construction~\ref{constr:strong} outputs a
  strongly $(d,e;u)$-disjunct matrix for every $d < (\tn -
  e)/((\tn-\td)u)$.
\end{lem}

\begin{Proof}
  Let $C := \{ c_1, \ldots, c_u \} \subseteq [n]$ and $C' := \{ c'_1,
  \ldots, c'_d \} \subseteq [n]$ be disjoint subsets of column
  indices. We wish to show that, for more than $e$ rows of $\cM$, the
  entries at positions picked by $C$ are all-ones while those picked
  by $C'$ are all-zeros. For each $j \in [n]$, denote the $j$th column
  of $\cM'$ by $\cM'(j)$, and let $\cM'(C) := \{ \cM'(c_j)\colon j \in
  [u] \}$, and $\cM'(C') := \{ \cM'(c'_j)\colon j \in [d] \}$.
 
  From the minimum distance of $\C$, we know that every two distinct
  columns of $\cM'$ agree in at most $\tn - \td$ positions.  By a
  union bound, for each $i \in [d]$, the number of positions where
  $\cM'(c'_i)$ agrees with one or more of the codewords in $\cM'(C)$
  is at most $u(\tn - \td)$, and the number of positions where some
  vector in $\cM'(C')$ agrees with one or more of those in $\cM'(C)$
  is at most $du(\tn - \td)$. 
  
  By assumption, we have $\tn - du(\tn -
  \td) > e$, and thus, for a set $E \subseteq [\tn]$ of size greater
  than $e$, at positions picked by $E$ none of the codewords in
  $\cM'(C')$ agree with any of the codewords in $\cM'(C)$.

  Now let $w \in [q]^n$ be any of the rows of $\cM'$ picked by $E$,
  and consider the $q^u \times n$ Boolean matrix $W$ formed by
  applying the mapping $\varphi(\cdot)$ on each entry of $w$. We know
  that $\{ w(c_j)\colon j \in [u]\} \cap \{ w(c'_j)\colon j \in [d]\}
  = \emptyset$.  Thus we observe that the particular row of $W$
  indexed by $(w(c_1), \ldots, w(c_u))$ (and in fact, any of its
  permutations) must have all-ones at positions picked by $C$ and
  all-zeros at those picked by $C'$. As any such row is a distinct row
  of $\cM$, it follows that $\cM$ is strongly $(d,e;u)$-disjunct.
\end{Proof}

Here we mention a few specific instantiations of the above
construction.  Namely, we will first consider the family of Reed-Solomon
codes, that are also used in the original work of Kautz and
Singleton~\cite{ref:KS64}, and then move on to the family of algebraic
geometric (AG) codes on the the Tsfasman-Vl{\u a}du{\c t}-Zink (TVZ)
bound, Hermitian codes, and finally, codes on the
Gilbert-Varshamov (GV) bound.

\subsection{Reed-Solomon codes} Let $p \in [0,1)$ be an arbitrary ``noise''
parameter.  If we take $\C$ to be an $[\tn, \tk, \td]_{\tn}$
Reed-Solomon code over an alphabet of size $\tn$ (which we assume
to be a prime power), where $\td =
\tn-\tk+1$, we get a strongly disjunct $(d,e;u)$-matrix with $m = O(du
\log n / (1-p))^{u+1}$ rows and $e = p \tn = \Omega(p d u (\log
n)/(1-p))$.

\subsection{Algebraic geometric codes on the TVZ bound} 
Another interesting family for the
code $\C$ is the family of algebraic geometric codes that attain the
Tsfasman-Vl{\u a}du{\c t}-Zink bound (cf.\ \cite{tsvz:82,gast:95}).
This family is defined over any alphabet size $q \geq 49$ that is a
square prime power, and achieves a minimum distance $\td \geq
\tn-\tk-\tn/(\sqrt{q}-1)$. Let $e := pn$, for a noise parameter $p \in
[0,1)$. By Lemma~\ref{lem:ks}, the underlying code $\C$ needs to have
minimum distance at least $\tn(1-(1-p)/(du))$. Thus in order to be
able to use the above-mentioned family of AG codes, we need to have $q
\gg (du/(1-p))^2 =: q_0$. Let us take an appropriate $q \in [2q_0,
8q_0]$, and following Lemma~\ref{lem:ks}, $\tn - \td = \lceil
\tn(1-p)/(du) \rceil$.  Thus, the dimension of $\C$ becomes at least
\[
\tk \geq \tn - \td - \frac{\tn}{\sqrt{q}-1} = \Omega \left(
  \frac{\tn(1-p)}{du} \right) = \Omega(\tn / \sqrt{q_0}),
\]
and subsequently\footnote{Note that, given the parameters $p, d, n$,
  the choice of $q$ depends on $p, d$, as explained above, and then
  one can choose the code length $\tn$ to be the smallest integer for
  which we have $q^\tk \geq n$. But for the sake of clarity we have
  assumed that $q^\tk = n$, which does not affect the asymptotic bounds.} 
  we get that $ \log n = \tk \log q \geq
\tk = \Omega(\tn / \sqrt{q_0}).  $ Now, noting that $m = q^u \tn$, we
conclude that
\[
m = q^u \tn = O(q_0^{u+1/2} \log n) = O\left(\frac{du}{1-p}
\right)^{2u+1} \log n,
\]
and $e = \Omega(p d u (\log n)/(1-p))$.

We see that the dependence of the number of measurements on the
sparsity parameter $d$ is worse for AG codes than Reed-Solomon codes
by a factor $d^u$, but the construction from AG codes benefits from a
linear dependence on $\log n$, compared to $\log^{u+1} n$ for
Reed-Solomon codes.  Thus, AG codes become more favorable only when the
sparsity is substantially low; namely, when $d \ll \log n$.

\subsection{Hermitian codes} A particularly nice family of AG codes arises
from the Hermitian function field.  Let $q'$ be a prime power and $q
:= q'^2$. Then the Hermitian function field over $\F_q$ is a finite
extension of the rational function field $\F_q(x)$, denoted by
$\F_q(x,y)$, where we have $y^{q'} + y = x^{q'+1}$.  The structure of
this function field is relatively well understood and the family of
Goppa codes defined over the rational points of the Hermitian function
field is known as Hermitian codes. This family is recently used by
Ben-Aroya and Ta-Shma \cite{ref:BT09} for construction of small-bias
sets. Below we quote some parameters of Hermitian codes from
their work.

\newcommand{\tg}{\tilde{g}}
The number of rational points of the Hermitian function field is equal
to ${q'}^3+1$, which includes a common pole $Q_\infty$ of $x$ and
$y$. The genus of the function field is $\tg = q'(q'-1)/2$. For some
integer parameter $r$, we take $G := rQ_\infty$ as the divisor
defining the Riemann-Roch space $\cL(G)$ of the code $\C$, and the set
of rational points except $Q_\infty$ as the evaluation points of the
code. Thus the length of $\C$ becomes $\tn = {q'}^3$. Moreover, the
minimum distance of the code is $\td = n-\deg(G) = n-r$. When $r \geq
2\tg - 1$, the dimension of the code is given by the Riemann-Roch
theorem, which is equal to $r-\tg+1$. For the low-degree regime where $r
< 2\tg-1$, the dimension $\tk$ of the code is the size of the
Wirestrauss semigroup of $G$, which turns out to be the set $W=\{
(i,j) \in \N^2\colon j \leq q'-1 \land iq'+j(q'+1) \leq r\}$.

Now, given parameters $d, p$ of the disjunct matrix, define $\rho :=
(1-p)/((d+1)u)$,
take the alphabet size $q$ as a square prime power,
and set $r := \rho q^{3/2}$. First we consider the case where $r < 2\tg
- 1 = 2q - 2\sqrt{q} - 1$.  In this case, the dimension of the
Hermitian code becomes $k = |W| = \Omega(r^2/q) = \Omega(\rho^2 q^2).$
The distance $\td$ of the code satisfies $\td = \tn - r \geq \tn
(1-\rho)$ and thus, for $e := p \tn$, conditions of Lemma~\ref{lem:ks}
are satisfied.  The number of the rows of the resulting measurement matrix
becomes $m = q^{u+3/2}$, and we have $n = q^\tk$. Therefore,
\[
\log n = k \log q \geq k = \Omega(\rho^2 q^2) \Rightarrow q =
O(\sqrt{\log n}/\rho) \Rightarrow m = O\left(\big(\frac{d\sqrt{\log
      n}}{1-p}\big)^{u+3/2}\right),
\]
and in order to ensure that $r < 2\tg-1$, we need to have $du/(1-p) \gg
\sqrt{\log n}$.  On the other hand, when $du/(1-p) \ll \sqrt{\log n}$,
we are in the high-degree regime, in which case the dimension of the
code becomes $k = r - \tg+1 = \Omega(r) = \Omega(\rho q^{3/2})$, and we
will thus have
\[
q = O((\log n / \rho)^{2/3}) \Rightarrow m = O\left(\big(\frac{d \log
    n}{1-p}\big)^{1+2u/3}\right)
\]
Altogether, we conclude that Construction~\ref{constr:strong} with
Hermitian codes results in a strongly $(d,e;u)$-disjunct matrix with
\[
m = O\left(\big( \frac{d \sqrt{\log n}}{1-p} + \big(\frac{d \log
    n}{1-p}\big)^{2/3} \big)^{u+3/2}\right)
\]
rows, where $e = p \cdot \Omega\left( d (\log n)/(1-p) + (d \sqrt{\log
    n} / (1-p))^{3/2} \right)$.  Compared to the Reed-Solomon codes,
the number of measurements has a slightly worse dependence on $d$, but
a much better dependence on $n$. Compared to AG codes on the TVZ
bound, the dependence on $d$ is better while the dependence on $n$ is
inferior.

\subsection{Codes on the Gilbert-Varshamov bound}
A $q$-ary $(\tn,\tk,\td)$-code (of sufficiently large length) is said
to be on the Gilbert-Varshamov bound if it satisfies $\tk \geq
\tn(1-h_q(\td/\tn))$, where $h_q(\cdot)$ is the $q$-ary entropy
function defined as
\[
h_q(x) := x \log_q(q-1) - x \log_q(x) - (1-x) \log_q(1-x).
\]
It is well known that a random linear code achieves the bound with
overwhelming probability (cf.\ \cite{ref:MS}).  Now we apply
Lemma~\ref{lem:ks} on a code on the GV bound, and calculate the
resulting parameters.  Let $\rho := (1-p)/(4du)$, choose any alphabet
size $q \in [1/\rho, 2/\rho]$, and let $\C$ be any $q$-ary code of
length $\tn$ on the GV bound, with minimum distance $\td \geq \tn
(1-2/q)$. By the Taylor expansion of the function $h_q(x)$ around $x =
1-1/q$, we see that the dimension of $\C$ asymptotically behaves
as $ \tk = \Theta(\tn / (q \log q)).  $ Thus, the number of columns of
the resulting measurement matrix becomes $n = q^\tk = 2^{\Omega(\tn /
  q)}$. Moreover, the number $m$ of its rows becomes
\[
m = q^u \tn = O(q^{u+1} \log n) = O((d/(1-p))^{u+1} \log n),
\]
and the matrix becomes strongly $(d,e;u)$-disjunct for $e = p \tn =
\Omega(p d (\log n)/(1-p))$.

We remark that for the range of parameters that we are interested in,
Porat and Rothschild \cite{ref:PR08} have obtained a
deterministic construction of linear codes on the GV bound that runs
in time $\poly(q^\tk)$ (and thus, polynomial in the size of the
resulting measurement matrix).

Their construction is based on a
derandomization of the probabilistic argument for random linear codes
using the method of conditional expectations, and as such, can be
considered \emph{weakly explicit} (in the sense that, the entire
measurement matrix can be computed in polynomial time in its length;
whereas for a fully explicit construction one must ideally be able to
deterministically compute any single entry of the measurement matrix
in time $\poly(d, \log n)$, which is not the case for this
construction). Altogether, we obtain the following result.

\begin{thm} \label{thm:GVdisjunct}
There is an algorithm that, given integer parameters $d \leq n$ and $u > 0$
and real parameter $p \in [0,1)$
outputs an $m \times n$ binary matrix which is strongly $(d,e;u)$-disjunct.
The parameters $m$ and $e$ satisfy the bounds 
$m=O((d/(1-p))^{u+1} \log n$ and 
$e=\Omega(p d (\log n)/(1-p))$. Moreover, the running time
of the algorithm is polynomial in $mn$. \qed
\end{thm}

Using a standard probabilistic argument it is easy to see that a
random $m \times n$ matrix, where each entry is an independent
Bernoulli random variable with probability $1/d$ of being $1$, is with
overwhelming probability strongly $(d,e;u)$-disjunct for $e = \Omega(p
d \log (n/d)/(1-p)^2)$ and $m = O(d^{u+1} (\log (n/d))/(1-p)^2)$ (the
proof is very similar to the proof of Lemma~\ref{lem:probDisjunct}).
Thus we see that, for a fixed $p$, Construction~\ref{constr:strong}
when using codes on the GV bound almost matches these
parameters. Moreover, the explicit construction based on Reed-Solomon
codes possesses the ``right'' dependence on the sparsity $d$, while AG codes
on the TVZ bound have a matching dependence on the vector length $n$
with random measurement matrices, and finally, the trade-off offered
by the construction based on Hermitian codes lies in between the one
for Reed-Solomon codes and AG codes.

\section{Concluding remarks} \label{sec:concl}

In this work we have introduced the combinatorial notion of
regular binary matrices, that is used as an intermediate tool towards
obtaining threshold testing designs. 

Even though our construction,
assuming an optimal lossless condenser, matches the probabilistic
upper bound for regular matrices, the number of measurements in the
resulting threshold testing scheme (obtained from the simple direct
product in Construction~\ref{constr:replace}) becomes larger than the
probabilistic upper bound by a factor of $\Omega(d \log n)$.  Thus, an
outstanding question is directly constructing threshold
disjunct matrices that match the probabilistic upper bound.  Despite this,
the notion of regular matrices may be of independent interest, and an
interesting question is to obtain (nontrivial) concrete lower bounds on the
number of rows of such matrices in terms of the parameters $n, d, e, u$ 
(and the gap parameter $g$ in the generalized definition of Section~\ref{sec:gap}).

Moreover, in this work we have assumed the upper threshold $u$ to be a fixed
constant, allowing the constants hidden in asymptotic notions
to have a poor dependence on $u$.  An outstanding question is whether
the number of measurements can be reasonably controlled when the upper threshold $u$
and possibly the gap parameter $g$ become large; e.g., $g, u = \Omega(d)$. 

The lower bound proved in Corollary~\ref{coro:lowerboundGeneralSimplified} on
the number of rows of threshold designs shows an exponent $g+2$ for the sparsity
parameter, which matches the upper bounds obtained using the probabilistic method.
We conjecture that
this bound can be improved to $\Omega_{u}(d^{g+2} \log_d n)$ and more generally
when $e>2$, to $\Omega_{u}(d^{g+2} \log_d n + d^{g+1} e)$. In other words,
for fixed thresholds, we suspect that the asymptotic bounds for $(d,e;u,g)$-threshold designs
and strongly $(d,e;g+1)$-disjunct matrices should nearly be the same.

Another interesting problem is
decoding. While our constructions can combinatorially guarantee 
identification of sparse vectors, for applications it is important to have
an efficient reconstruction algorithm as well. Contrary to the case of
strongly disjunct matrices that allow a straightforward decoding
procedure (cf. \cite{ref:thresh2}), it is not clear whether in general
our notion of disjunct matrices allow efficient decoding, and thus it becomes
important to look for constructions that are equipped with efficient
reconstruction algorithms. 

Finally, for clarity of the exposition, in
this work we have only focused on asymptotic trade-offs, and it would
be nice to obtain good, finite length, estimates on the obtained bounds 
that are useful for applications.

%
%


\section*{Acknowledgments}

The author thanks Christian Deppe, Arkadii D'yachkov, and Vyacheslav Rykov
for fruitful discussions, and anonymous referees for their feedback on
earlier drafts of this work.

\bibliography{threshold}

\begin{appendix}

%





\section{A technical lemma} \label{app:lemma}

The following simple proposition is used in the proof of Theorem~\ref{thm:lowerboundGeneral}.

\begin{prop} \label{prop:technical}
Suppose for some values of $a > 0$, $b,m \geq 2$, and $c \geq 2b/a$, we have
$m \geq c \cdot \frac{a}{b+\log m}$, where the logarithm is to base $2$.
Then,
\[
m \geq \frac{(ac/b)}{\log(ac/b)}.
\]
\end{prop}

\begin{proof}
We can write
\[
m \geq c \cdot \frac{a}{b+\log m} \geq \frac{ca}{b \log m} \Rightarrow
m \log m \geq ac/b,
\]
where the second inequality is from the assumption that $b, m \geq 2$.

Since $m \log m$ is an increasing and convex function of $m$, 
we know that $m \geq m_0$, where $m_0$ is the solution to the equation
$m_0 \log m_0 = ac/b$. Thus it suffices to lower bound $m_0$.

Since $ac/b \geq 2$ by assumption, it follows that $m_0 \leq m_0 \log m_0 = ac/b$,
and thus,
\[
m_0 \log(ac/b) \geq m_0 \log m_0 = ac/b \Rightarrow m_0 \geq \frac{(ac/b)}{\log(ac/b)},
\]
and the claim follows.
\end{proof}

\end{appendix}

\end{document}